\documentclass{aa}  
\usepackage[varg]{txfonts} 
\usepackage{graphicx}   
\usepackage[hypertexnames=false]{hyperref}
\usepackage{amsmath}    
\usepackage{amssymb}
\usepackage{array}
\usepackage{siunitx}
\usepackage{dcolumn}
\usepackage{pdfpages}
\usepackage{multirow}
\usepackage{threeparttable}
\usepackage{placeins}
\usepackage{float}

\newcommand{\kms}{km\,s$^{-1}$}                             
\newcommand{\mum}{$\mu$m}                                           
\newcommand{\msun}{M$_{\sun}$}                  
\newcommand{\lsun}{L$_{\sun}$}                  
\newcommand{\rsun}{R$_{\sun}$}                  
\def\H2{H$_2$}                                  
\newcommand{\HII}{\ion{H}{II}}     
\newcommand{\radec}{($\alpha,\delta$)(J2000)}   
\newcommand{\Spitzer}{{\slshape{Spitzer}}}          
\newcommand{\Herschel}{{\slshape{Herschel}}}    

\begin{document}

\title{High-resolution images of two wiggling stellar jets, MHO~1502\\ 
and MHO~2147, obtained with \texorpdfstring{GSAOI$+$GeMS}{}}

\titlerunning{High-resolution images of the stellar jets MHO~1502 and MHO~2147 obtained with \texorpdfstring{GSAOI$+$GeMS}{}}

\author{L.V. Ferrero \inst{1,2}
\and G. G\"unthardt  \inst{1}
\and L. Garc\'ia \inst{1}
\and M. G\'omez \inst{1,2}
\and V.M. Kalari \inst{3,4}
\and H.P. Salda\~no \inst{1}
}

\institute{Universidad Nacional de C\'ordoba, Observatorio Astorn\'omico de C\'ordoba, Laprida 854, X5000BGR C\'ordoba, Argentina. 
\and
Consejo Nacional de Investigaciones Cient\'ificas y T\'ecnicas (CONICET), Godoy Cruz 2290, CABA, CPC 1425FQB, Argentina. 
\and
Departamento de Astronom\'ia, Universidad de Chile, Casilla 36-D, Santiago, Chile.
\and
Gemini Observatory/NSF’s NOIRLab, Casilla 603, La Serena, Chile.\\ 
\email{lvferrero@unc.edu.ar}
}

\date{Received ; accepted }


\abstract
{}
{We investigated the possible cause--effect relation between the wiggling shape of two stellar jets, MHO~1502 and MHO~2147, and the potential binarity of the respective driving stars.}
{We present high-angular-resolution \H2\ (2.122~\mum) and K-band images obtained with the Gemini South Adaptive Optics Imager (GSAOI) and the Gemini Multi-conjugate Adaptive Optics System (GeMS). The profiles of the jets are depicted in detail by the \H2\ images. We used K-band data to search for potential close companions to the previously suggested exciting sources, and used archive data to investigate these sources and the environments in which the jets are located. We also applied a model to reproduce the wiggling profiles of the jets.} 
{MHO~1502 is composed of a chain of knots delineating the wiggling jet, suggesting that the driving source emitted them in an intermittent manner. Our K-band image of the previously proposed exciting star, IRAC~18064, shows two sources separated by $\sim240$~AU, hinting at its binarity. However, as IRAC~18064 is located off the jet axis at $\sim2064$~AU, it is questionable as to whether this source is the true exciting star. Moreover, the orbital model centred on IRAC~18064 suggests a binary companion at a much greater distance ($\sim2200$~AU) than the nearby star (at $\sim$240~AU). On the other hand, the orbital model centred on the axis provides the best fits. Nevertheless, the precession model centred on the axis cannot be discarded, despite having larger residuals and $\chi^2$. MHO~2147 displays an S-shaped gentle continuous emission in \H2. We identify two other jets in the field of MHO~2147: a previously reported quasi-perpendicular jet, MHO~2148, and a third jet adjacent to MHO~2147. The model that best fits the morphology of the MHO~2147 jet and that of its adjacent jet is precession. The exciting source of MHO~2147 may be a triple system.}
{}


\maketitle
\section{Introduction}

Stellar jets and molecular outflows are ubiquitous in star-forming regions to such an extent that they are considered to be signposts of newly formed stars. Furthermore, they are usually associated with the earliest stages of stellar formation, when the incipient proto-stellar object is deeply embedded within a dense core of dust and molecular gas \citep{Reipurth2000,Reipurth-Bally2001,Frank2014}. In the optical and near-infrared ranges, jets usually consist of a chain of knots with a terminal leading bow shock emanating from the young star and propagating away at high supersonic speeds. Although most jets are straight or collimated, others have curved shapes. Different mechanisms can produce the non-straight jets, such as the interstellar magnetic field, which curves or bends the jet axis \citep{Fendt-Zinnecker1998}. In addition, the stellar magnetosphere \citep{Lii2014, Dyda2015} or magnetized disc can yield asymmetries in the jet \citep{Fendt-Sheikhnezami2013}.  In other cases, the wiggling profile may indicate that the jet axis wanders (precesses), or that the exciting source of the jet undergoes orbital motion.

The shape of a non-straight or wandering jet axis may be related to some characteristics of the central object. For example, reflection-symmetric wiggles may be due to the orbital motion of the jet source that belongs to a binary system, whereas the precession of the jet axis could produce a point-symmetric S-shaped jet induced by a tidal effect caused by a companion star in an orbit not coplanar to the disc\footnote{Asymmetries in the circumstellar disc may also provoke a tidal effect on the jet source.} \citep{Anglada2007,Lee2010,Estalella2012,Sanchez-Monge2014,Beltran2016,Paron2016}. Nevertheless, in many cases, because of the obscuration towards the exciting source and/or the limited angular resolution of the observations, the driving star cannot be resolved. However, the large-scale morphology of a jet, in combination with a simple model, can provide a means to extrapolate backwards and infer some properties of the exciting star(s) and the jet. To unveil faint substructures and reveal the complex morphologies of these jets, high-spatial-resolution and high signal-to-noise-ratio data are needed. Near-infrared observations, and in particular the 2.12~\mum\ \H2 line, which is a well-known tracer of shock-excited regions \citep{Wakelam2017}, provide a key tool for gaining insights into the relation between the jet and the forming central star.

The number of mirror-symmetric and S-shape jets detected has been increasing steadily during recent decades, even though they still represent a small fraction of known stellar jets, and in particular, of the \H2 molecular jets. In addition, only a few have been imaged with sufficiently high spatial resolution and sensitivity to be able to detect small-scale structures and to map their convoluted shapes \citep{Davis2010}.

With the aim of investigating the likely cause--effect relation between the wiggling morphology of stellar jets and the central star properties, in this study, we present high-resolution \H2 (2.122~\mum) and K-band images obtained with the Gemini South Adaptive Optics Imager (GSAOI) and the Gemini Multi-conjugate Adaptive Optics System (GeMS) of \object{MHO~1502} and \object{MHO~2147}. 

Previous low-resolution \H2 and 4.5~\mum\ IRAC/\Spitzer\ images of MHO~1502 revealed a chain of knots delineating a meandering profile \citep{DeLuca2007,Giannini2007}. \cite{Giannini2013} suggested that \object{[SEC2010]~IRAC~18064} (hereafter \object{IRAC~18064}) is the driving source, despite it being located $\sim 0.01$~pc away from the axis of the jet. Meanwhile, MHO~2147 displays a gentle, S-shaped continuous emission of \H2.  \cite{Varricatt2011} suggested that the likely driving source \object{IRAS~17527$-$2439} (hereafter IRAS~17527) is a deeply embedded Class~I protostar. This author also reported the detection of a significantly fainter and almost perpendicular jet to MHO~2147, designated  \object{MHO~2148}.

We investigated the surrounding regions of the jets and their relation with the potential driving sources. In particular, we searched for the binarity or multiplicity of the exciting stars in our high-resolution K-band images. We also applied a simple model developed by \cite{Masciadri2002} to reproduce the wiggling profiles, and investigated the proposed exciting source for each jet.

A description of the observations and the instrument employed are presented in \S \ref{sec_observations}. In \S \ref{sec_analysis}, we investigate the morphology and internal structure of each jet as well as their environments. We also analyse the probable multiplicity of the exciting sources. We reproduced the jet profiles of MHO~1502 and MHO~2147, and that of the jet adjacent to MHO~2147,  using a model to estimate the jets and the exciting source parameters in \S \ref{sec_models_masciadri}. Finally, we discuss our results and present our conclusions (see \S \ref{sec_conclusions}).

\section{Observations}
\label{sec_observations}

\subsection{\texorpdfstring{GSAOI$+$GeMS}{} data}
    
We used the GSAOI instrument and the GeMS mounted at the 8.1m Gemini South Telescope in Cerro Pachon \citep{McGregor2004,Carrasco2012}. GeMS is an adaptive optics (AO) system that uses five sodium Laser Guide Stars (LSG) with up to three Natural Guide Stars (NGSs) and multiple deformable mirrors (DMs) optically conjugated with the main turbulence layers. This provides an AO-corrected field that is larger than a single-conjugated AO  \citep[AO-SCAO;][]{Neichel2014a,Rigaut2014}. GSAOI$+$GeMS provides diffraction-limited images in the $0.9 - 2.4$~\mum\ wavelength range over a $85''\times85''$ field of view (FoV), with an imaging scale of $0\farcs0197$\,pixel$^{-1}$. The GSAOI detector is composed of a $2\times2$ Rockwell HAWAII-2RG $2048\times2048$ pixel array mosaic with four gaps between the arrays of $\sim2.4$~mm, corresponding to $\sim2.5''$ on the sky.

The images for MHO~1502 and MHO~2147/2148 obtained using the \H2 (1-0 S(1), \hbox{$\lambda_{\rm c}= 2.122$}~\mum, \hbox{$\Delta\lambda = 0.032$}~\mum) and K (\hbox{$\lambda_{\rm c}= 2.200$}~\mum, \hbox{$\Delta\lambda = 0.34$}~\mum) filters were taken between February 11 and May 27, 2014 (Program ID: GS-2014A-Q-29). To remove the gaps between the detectors, we used a $3\times3$ dither pattern with steps of $5''$ and $7''$ for MHO~1502 and MHO~2147/2148, respectively. In the case of MHO~1502, we obtained a total of 10 science fields, with individual exposure times of 200~s each for the \H2 filter, and 9 science fields of 49~s each for the K-filter. For MHO~2147, 11 and 9 science fields of 200~s (\H2 filter) and 60~s (K-band filter), respectively, were observed. The data were processed and combined to obtain the final mosaic for each filter with THELI\footnote{THELI is a tool for the automated reduction of astronomical images in optical, near- and mid-infrared, available at the website: \url{https://www.astro.uni-bonn.de/theli/}} \citep{Schirmer2013,Erben2005}. The reduction process was similar to that described in \cite{Schirmer2013} and \cite{Schirmer2015}. The science frames were also used for removing the sky background contribution. 

 To obtain the astrometry and the distortion correction in the K-band, the 2MASS catalogue was used for both the MHO~1502 and MHO~2147/2148 fields. However, due to the small FoV of GSAOI, the number of  stars in the all-sky astrometric reference 2MASS catalogue was insufficient for the calibration of the individual frames. Thus, secondary reference catalogues were needed. For MHO~2147, the VISTA\footnote{VISTA Science Archive: \url{http://horus.roe.ac.uk/vsa/index.html}} catalogue was used. In the case of MHO~1502, images from the VLT Infrared Spectrometer And Array Camera (ISAAC\footnote{ESO Archive: \url{http://archive.eso.org/cms/eso-data.html}}) were employed. We estimated astrometric uncertainties of 0.28\arcsec\ and 0.19\arcsec\ for the MHO~1502 and the MHO~2147 fields, respectively.

In the case of MHO~1502, we detected an elongation effect ($e\approx0.1$) on the sources in the upper left corner of the field. K and \H2 images of this object were taken on February 11, 2014. The elongation effect was likely produced by a fault in the lasers of the AO system, as was reported when the data were collected\footnote{Reported in the `Status and Availability' of GSAOI on February 21, 2014: \url{https://www.gemini.edu/sciops/instruments/gsaoi/status-and-availability}}. This failure was fixed in time for our April run, and no elongation effect was seen in the field of MHO~2147, whose images were obtained on May 27, 2014, for the K-band data and on April 16 and May 27, 2014, for the \H2 images. The final images had FWHMs that varied over the FoV by about 8\% for MHO~1502, and 5\% for MHO~2147. For MHO~1502, the FWHM was of 0.11\arcsec\ and 0.14\arcsec\ (77 and 98~AU, at a distance of 700~pc) for the \H2 and K-band, respectively. For MHO~2147/2148, the FWHM was 0.09\arcsec\ for \H2 images and 0.08\arcsec\ for K images (290 and 258~AU, at a distance of 3.23~Kpc). The resolution achieved in each case is in agreement with the values reported by \cite{Neichel2014a}. To improve the signal-to-noise ratio of the faint and diffuse structures, the reduced mosaic images were convolved with a Gaussian filter of 2~pixels (0.22\arcsec\ for MHO~1502 and 0.18\arcsec\ for MHO~2147). 

Figures \ref{fig_MHO1502} and \ref{fig_MHO2147} show our mosaic images for MHO~1502 and MHO~2147, which are discussed in Sect.~\ref{sec_analysis}. The nomenclature that we adopt for the knots is the same as that used in classical stellar jets, where a capital letter represents a knot and the number following this letter denotes a substructure within the knot. Enlarged images of each knot (Figs.~\ref{fig_MHO1502_ABCDEF} and \ref{fig_MHO1502_GHIJ} for MHO~1502, Figs.~\ref{fig_MHO2147_ABCDEFG} and \ref{fig_MHO2147_HI} for MHO~2147, and Fig.~\ref{fig_MHO2148_AB} for MHO~2148) reveal their internal structure in detail. The white arrows in the enlarged figures mark the centre of the knots listed in Tables~\ref{tab_photo1502} and \ref{tab_photo2147}.

We used the aperture photometry package APPHOT in IRAF\footnote{IRAF is distributed by the National Optical Astronomy Observatory, which is operated by the Association of Universities for Research in Astronomy, Inc., under cooperative agreement with the National Science Foundation.} to estimate the flux of each knot by subtracting the sky background images from the science images and calibrating the fluxes taking into account that the width of the \H2-filter is about ten times smaller than the width of the K-band filter. 

The 2MASS K$_s$ band magnitudes of several field stars were employed to flux calibrate the \H2~(1-0) S(1) images. In the case of MHO~1502, ten 2MASS stars were used, while for MHO~2147, it was possible to use only four stars due to saturation of bright 2MASS stars. The fluxes were measured taking into account a 3-$\sigma$ threshold between the background rms and \H2 emission. The coordinates of the knots and final fluxes are listed in Tables~\ref{tab_photo1502} and \ref{tab_photo2147}, where the $r$ parameter denotes the radius of the circular aperture used for the photometry. There are special cases where an ellipse instead of a circle was used to calculate the photometry because of the irregular and elongated morphology of some knots (see Tables \ref{tab_photo1502} and \ref{tab_photo2147}, Col. 6). To estimate the errors, the uncertainties resulting from the conversion factor derived from the 2MASS K$_s$ magnitudes and the sky background variations in the \H2 filter were considered.

\subsection{Complementary data}

To complement our near-infrared data, we used four-band IRAC images centred at 3.6, 4.5, 5.8, 8.0~\mum, as well as 24\,\mum\ MIPS images obtained from the Galactic Legacy Infrared Mid-Plane Survey Extraordinaire (GLIMPSE; \citealt{Benjamin2003}) and the \Spitzer\ Enhanced Imaging Products\footnote{Available at \url{http://irsa.ipac.caltech.edu/data/SPITZER/Enhanced/SEIP/}}. We also analysed \Herschel\ data in 70 and 160~\mum\ images taken from the \Herschel-PACS Point Source Catalogue \citep[HPPSC,][]{Marton2017, Herschel_PACS_Vizier_2020}, in addition to 250, 350, and 500~\mum\ images obtained from the \Herschel\ Infrared GAlactic Plane Survey \citep[Hi-GAL,][]{Molinari2010}. Finally, we made use of data at 870~\mum, retrieved from the APEX Telescope LArge Survey of the GALaxy\footnote{\url{http://atlasgal.mpifr-bonn.mpg.de/cgi-bin/ATLASGAL_DATABASE.cgi}} \citep[ATLASGAL,][]{Schuller2009}.
    
\section{The MHO~1502 and MHO~2147 jets}
\label{sec_analysis}

\begin{figure*}
  \centering
  \includegraphics[width=\textwidth]{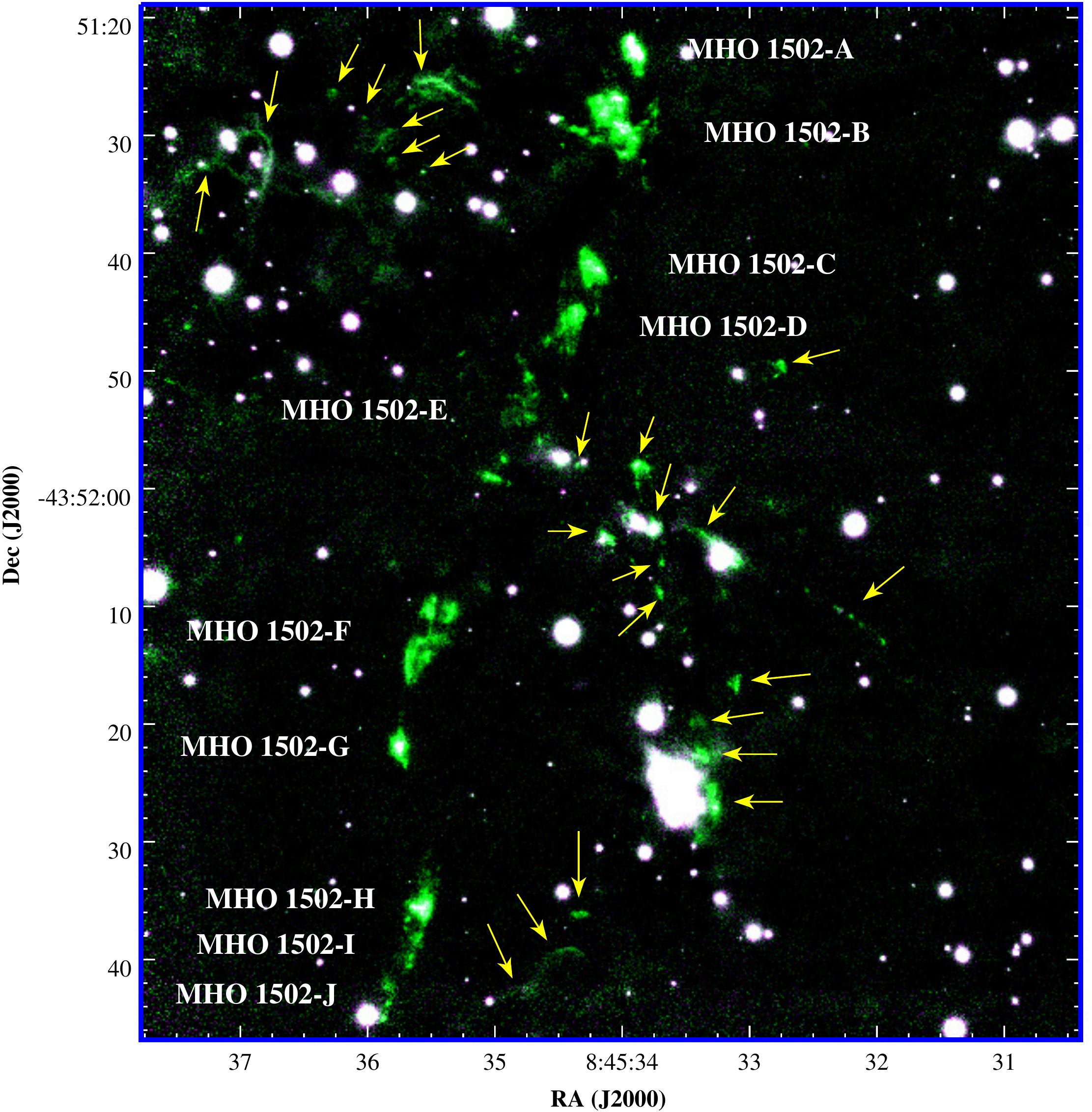}
  \caption{Composite image of MHO~1502 obtained with GSAOI/GEMINI. The K-band filter is shown in magenta and the \H2-band filter in green. The yellow arrows indicate \H2 emission adjacent to the MHO~1502 jet, which lie in the field and are unlikely to be associated with this jet (see Sect.~\ref{sec_adj1502}).}
  \label{fig_MHO1502}
\end{figure*}

\subsection{MHO~1502} 

\subsubsection{Characteristics of the jet}
\label{sec_MHO1502}

MHO~1502 is located in the molecular cloud Vela-D at a distance of 700~pc \citep{Liseau1992}, in the vicinity of the \HII\ region \object{G263.619$-$0.53} and the young stellar cluster IRS~16 \citep{Caswell-Haynes1987,Massi2010}. This object was discovered in \H2 by \cite{DeLuca2007} and was observed in the 4.5~\mum\ IRAC/\Spitzer\ band by \cite{Giannini2007} and catalogued by \cite{Davis2010}. \cite{Giannini2013} identified the following eight knots along the wiggling jet axis (see Fig.~4 in their work): Knots 1-2-3-4, corresponding to the NW blueshifted component, and knots 5-6-7-8 belonging to the SE redshifted lobe.

\cite{Massi2007} identified a dust continuum core at 1.2~mm MMS2 near to the centre of the jet but not exactly coincident with it. \cite{Giannini2013} suggested that the source IRAC~18064, located close to the centre of the jet, but again not perfectly aligned with it, may be the exciting source. These authors combined fluxes from 3.6~\mum\ to 1.2~mm to construct and model the SED, and obtained a Class~I spectral index and derived several parameters: in particular, an envelope mass of 43~\msun\ and a central object mass of 0.87~\msun. Considering the amount of mass in the stellar embryo plus the envelope, these latter authors suggested that the driving source could be an intermediate-mass object, although an unresolved binary or multiple system could not be excluded (in the IRAC bands). Our K-band GSAOI/Gemini image (see Fig.~\ref{fig_sources_k_filter}, left panel) resolves IRAC~18064 into two components, with a projected angular separation of 0.3\arcsec\ (or 240~AU, at a distance of 700~pc), suggesting binarity of this proposed driving source.

Figure~\ref{fig_MHO1502} shows our GSAOI/Gemini combined \H2 (in green) and K (in magenta) image of MHO~1502, with the gentle wiggling profile composed of a chain of consecutive knots being shown in detail. The jet has a total extension of 87\arcsec\ (0.3~pc, assuming a distance of 700~pc) from NW to SE and a position angle of $165\degr$. The eight knots catalogued by \cite{Giannini2013} were identified, as well as two others not previously reported (knots MHO~1502-I and J) located at the bottom left of the image. The knots were renamed using the convention explained in Sect.~\ref{sec_observations}. Table~\ref{tab_photo1502} lists the knots shown in Fig.~\ref{fig_MHO1502}. The high spatial resolution obtained with AOs (3-6 times better than other studies), makes it possible to recognise more details in each knot than in previous works. Figures~\ref{fig_MHO1502_ABCDEF} and \ref{fig_MHO1502_GHIJ} display enlarged \H2 flux-calibrated images of the NW and SE knots, respectively. In general, these knots presented irregular morphologies and complex structures with many sub-knots surrounded by fainter diffuse \H2 emission (see e.g. knots B and F in Fig.~\ref{fig_MHO1502_ABCDEF}, upper and lower right panels). In other cases, no central emission was detected in the knot, but diffuse and faint (nebula-like) structures were found (such as knots I and J in Fig.~\ref{fig_MHO1502_GHIJ}, lower panels).

In the \H2\ band image of the lower left panel of Fig.~\ref{fig_MHO1502_ABCDEF}, we detect three components plus a dim object to the northeast of IRAC~18064 \citep{Giannini2013}. On the other hand, in the K-band image of the left panel of Fig.~\ref{fig_sources_k_filter}, we detect two sources plus a very faint object (dim object) to the northeat. The most southeasterly source in Fig.~\ref{fig_MHO1502_ABCDEF}, labelled `new knot', is located at \radec~$=$~(08:45:34.51; $-$43:51:57.7) and is marked with a white arrow. As it has no counterpart in the K-band, this suggests that it is a shock emission object or a knot that has been recently ejected by the young star IRAC~18064. This new knot lies only 0.25\arcsec\ from the position of IRAC~18064. Assuming a velocity of 100~\kms \citep[see e.g.][]{Bally2016}, we determin that this knot was emitted only 8~years ago. However, the position of the new knot on the image is orientated almost perpendicular and displaced from the jet axis, suggesting that it is unlikely to be associated with MHO~1502.

\begin{table*}
 \centering
\begin{threeparttable}
\caption{Coordinates and \texorpdfstring{\H2}{} fluxes for the knots associated with the MHO~1502 jets.}
 \label{tab_photo1502} 
 \begin{tabular}{
        llcc
        S[table-format = 3.2]@{\,\( \pm \)\,}
        S[table-format = 1.2]
        S[table-format = 1.2]
        cl}
 \hline \hline
 \noalign{\smallskip}
  \multicolumn{2}{c}{\multirow{2}{4em}{Knot ID}}  &
  $\alpha$ (J2000.0)        & 
  $\delta$ (J2000.0)        & 
  \multicolumn{2}{c}{Flux}  &    
  {r\tablefootmark{a}}                       &    
  MHO ID\tablefootmark{b}   &
  \multirow{2}{10em}{Fig. Reference\tablefootmark{c}}  \\
        &                        & 
    ($^{h}:^{m}:^{s}$)                  &
    (\degr:\arcmin:\arcsec)             &
    \multicolumn{2}{c}{(10$^{-6}$ Jy)}  &
    {(\arcsec)}                         &
    {(jet-knot)}                        &             \\
  \hline
\noalign{\smallskip} 
            A & 1   & 8:45:33.94 & -43:51:21.8 & 12.61 &  0.92  & 0.29 &  1--1 & \multirow{4}{10em}{\ref{fig_MHO1502_ABCDEF}, upper left panel}  \\
              & 2   & 8:45:33.93 & -43:51:22.2 & 13.15 &  0.94  & 0.23 &  1--1&                             \\
              & 3   & 8:45:33.70 & -43:51:22.7 & 102.14 &  7.32 & 0.91 &  1--1&                             \\
              & 4\tablefootmark{d}  & 8:45:33.87 & -43:51:24.6 &      \multicolumn{2}{c}{$-$}          &  {$-$}  &  1--1&                        \\
\hline
\noalign{\smallskip}
            B & 1   & 8:45:34.18 & -43:51:26.9 & 7.09 &  0.51   & 0.30 &  1--2 & \multirow{17}{10em}{\ref{fig_MHO1502_ABCDEF}, upper right panel}    \\
              & 2   & 8:45:34.05 & -43:51:26.9 & 31.81 &  2.51  & 0.67\tablefootmark{e} &  1--2&          \\
              & 3   & 8:45:34.07 & -43:51:27.5 & 5.47 &  0.39   & 0.30  &  1--2&    \\
              & 4   & 8:45:34.03 & -43:51:27.7 & 5.87 &  0.44   & 0.28  &  1--2&    \\
              & 5   & 8:45:34.16 & -43:51:28.1  & 4.97 &  0.35   & 0.22 &  1--2&    \\
              & 6   & 8:45:34.01 & -43:51:29.3 & 55.38 &  4.15  & 0.91  &  1--2&    \\ 
              & 7   & 8:45:33.90 & -43:51:30.3 & 6.39 &  0.50   & 0.34  &  1--2&    \\
              & 8   & 8:45:33.88 & -43:51:29.4 & 2.05 &  0.17   & 0.24  &  1--2&    \\
              & 9   & 8:45:33.81 & -43:51:28.8 & 2.46 &  0.31   & 0.40  &  1--2&    \\
              & 10  & 8:45:33.74 & -43:51:28.0 & 3.24 &  0.32   & 0.41  &  1--2&    \\
              & 11  & 8:45:33.78 & -43:51:27.4 & 1.35 &  0.18   & 0.31  &  1--2&    \\
              & 12  & 8:45:34.49 & -43:51:28.5 & 2.75 &  0.23   & 0.30  &  1--2&    \\
              & 13  & 8:45:34.38 & -43:51:29.4 & 3.10 &  0.71   & 0.34  &  1--2&    \\
              & 14  & 8:45:34.31 & -43:51:29.6 & 4.65 &  0.34   & 0.37  &  1--2&    \\
              & 15  & 8:45:34.21 & -43:51:30.4 & 9.30 &  1.08   & 0.73  &  1--2&    \\
              & 16  & 8:45:33.99 & -43:51:31.5 & 8.30 &  0.61   & 0.39  &  1--2&    \\
              & 17  & 8:45:33.84 & -43:51:32.1 & 1.29 &  0.26   & 0.39  &  1--2&    \\              
\hline
\noalign{\smallskip}
            C & 1   & 8:45:34.32 & -43:51:39.9 & 1.71 &  0.14   & 0.20  &  1--3& \multirow{4}{10em}{\ref{fig_MHO1502_ABCDEF}, middle left panel}   \\
              & 2   & 8:45:34.27 & -43:51:40.7 & 6.17 &  0.47   & 0.29  &  1--3&   \\
              & 3   & 8:45:34.12 & -43:51:41.3 & 15.75 &  1.16  & 0.43  &  1--3&   \\
              & 4   & 8:45:34.27 & -43:51:42.7 & 1.95 &  0.21   & 0.22  &  1--3&   \\
\hline
\noalign{\smallskip}
            D & 1   & 8:45:34.37 & -43:51:45.2 & 2.52 &  0.19   & 0.17  &  1--4& \multirow{4}{10em}{\ref{fig_MHO1502_ABCDEF}, middle right panel}   \\ 
              & 2   & 8:45:34.47 & -43:51:45.7 & 3.51 &  0.43   & 0.33  &  1--4&    \\
              & 3   & 8:45:34.47 & -43:51:46.4 & 1.12 &  0.10   & 0.20           &  1--4&    \\
              & 4   & 8:45:34.59 & -43:51:47.4 & 4.22 &  0.70   & 0.63  &  1--4&    \\
\hline
\noalign{\smallskip}
            E & 1   & 8:45:34.75 & -43:51:50.5 & 4.26 &  0.98   & 0.61           &  1--4& \multirow{8}{10em}{\ref{fig_MHO1502_ABCDEF}, lower left panel}   \\ 
              & 2   & 8:45:34.46 & -43:51:52.0 & 1.25 & 0.13        & 0.26      &  1--4 &                           \\
              & 3\tablefootmark{d} & 8:45:34.78 & -43:51:53.3 &   \multicolumn{2}{c}{$-$}   & {$-$} &  1--4&   \\
              & 4   & 8:45:34.66 & -43:51:55.9 & 2.29 &  0.39   & 0.37  &  1--4&  \\
              & 5   & 8:45:34.94 & -43:51:57.5 & 2.48 &  0.50   & 0.37  &  1--4&    \\
              & 6a  & 8:45:35.06 & -43:51:58.9 & 3.72 &  0.30   & 0.27  &  1--5&      \\
              & 6b  & 8:45:35.00 & -43:51:59.0 & 1.72 &  0.12   & 0.18  &  1--5&      \\
\hline
\noalign{\smallskip}
            F & 1   & 8:45:35.48 & -43:52:09.4 & 5.11 &  0.55   & 0.45  &  1--6& \multirow{8}{10em}{\ref{fig_MHO1502_ABCDEF}, lower right panel}   \\
              & 2   & 8:45:35.53 & -43:52:09.9 & 15.51 &  1.20  & 0.61  & 1--6&   \\
              & 3   & 8:45:35.33 & -43:52:10.0 & 11.81 &  1.05  & 0.63  & 1--6&   \\
              & 4   & 8:45:35.37 & -43:52:10.9 & 2.32 &  0.19   & 0.24  &  1--6&    \\
              & 5   & 8:45:35.60 & -43:52:12.9 & 13.98 &  1.00  & 0.93\tablefootmark{e} &  1--6&    \\
              & 6   & 8:45:35.69 & -43:52:14.1 & 8.33 &  0.59   & 0.50  &  1--6&    \\
              & 7   & 8:45:35.69 & -43:52:15.1 & 5.98 &  0.43   & 0.38  &  1--6&    \\
              & 8   & 8:45:35.54 & -43:52:14.4 & 17.67 &  1.81  & 1.25\tablefootmark{e} &  1--6&    \\
\hline
\noalign{\smallskip}
            G &     & 8:45:35.76 & -43:52:21.8 & 53.81 &  4.26  & 0.89  &  1--7& \ref{fig_MHO1502_GHIJ}, upper left panel   \\
\hline
\noalign{\smallskip}
            H & 1   & 8:45:35.65 & -43:52:35.5 & 8.71 &  0.63   & 0.26      &  1--8   & \multirow{3}{10em}{\ref{fig_MHO1502_GHIJ}, upper right panel}   \\
              & 2   & 8:45:35.59 & -43:52:35.6 & 2.93 &  0.21   & 0.13  &  1--8&    \\
              & 3   & 8:45:35.58 & -43:52:36.0 & 1.81 &  0.13   & 0.11  &  1--8&    \\
\hline
\noalign{\smallskip}
            I & 1   & 8:45:35.61 & -43:52:38.1 & 8.16 &  0.86   & 0.83  &  {$-$}& \multirow{3}{10em}{\ref{fig_MHO1502_GHIJ}, lower left panel}  \\
              & 2   & 8:45:35.67 & -43:52:39.7 & 2.11 &  0.20   & 0.33  &  {$-$}&   \\
              & 3   & 8:45:35.68 & -43:52:40.4 & 2.12 &  0.16   & 0.33  &  {$-$}&   \\
\hline
\noalign{\smallskip}
            J & 1\tablefootmark{d}   & 8:45:35.86 & -43:52:42.3 &   \multicolumn{2}{c}{$-$} & {$-$} &  {$-$}& \multirow{3}{10em}{\ref{fig_MHO1502_GHIJ}, lower right panel}  \\
              & 2   & 8:45:35.84 & -43:52:43.7 & 1.96 &  0.50   & 0.50  &  {$-$}& \\
              & 3   & 8:45:35.87 & -43:52:44.9 & 5.78 &  0.61   & 0.53  &  {$-$}& \\
\hline
  \end{tabular}
 \tablefoot{
 \tablefoottext{a}{Radii of the circular apertures used for the photometry.}
 \tablefoottext{b}{MHO ID from \citet[Col. 7]{Giannini2013}.}
 \tablefoottext{c}{Reference to the figure and panel in which an enlarged version of each knot can be observed.}
 \tablefoottext{d}{Flux lower than 3$\sigma$.}
 \tablefoottext{e}{Photometry on an ellipse with semi-major axis r.}}
 \end{threeparttable}
\end{table*}

\begin{figure*} 
  \centering
  \includegraphics[width=\columnwidth]{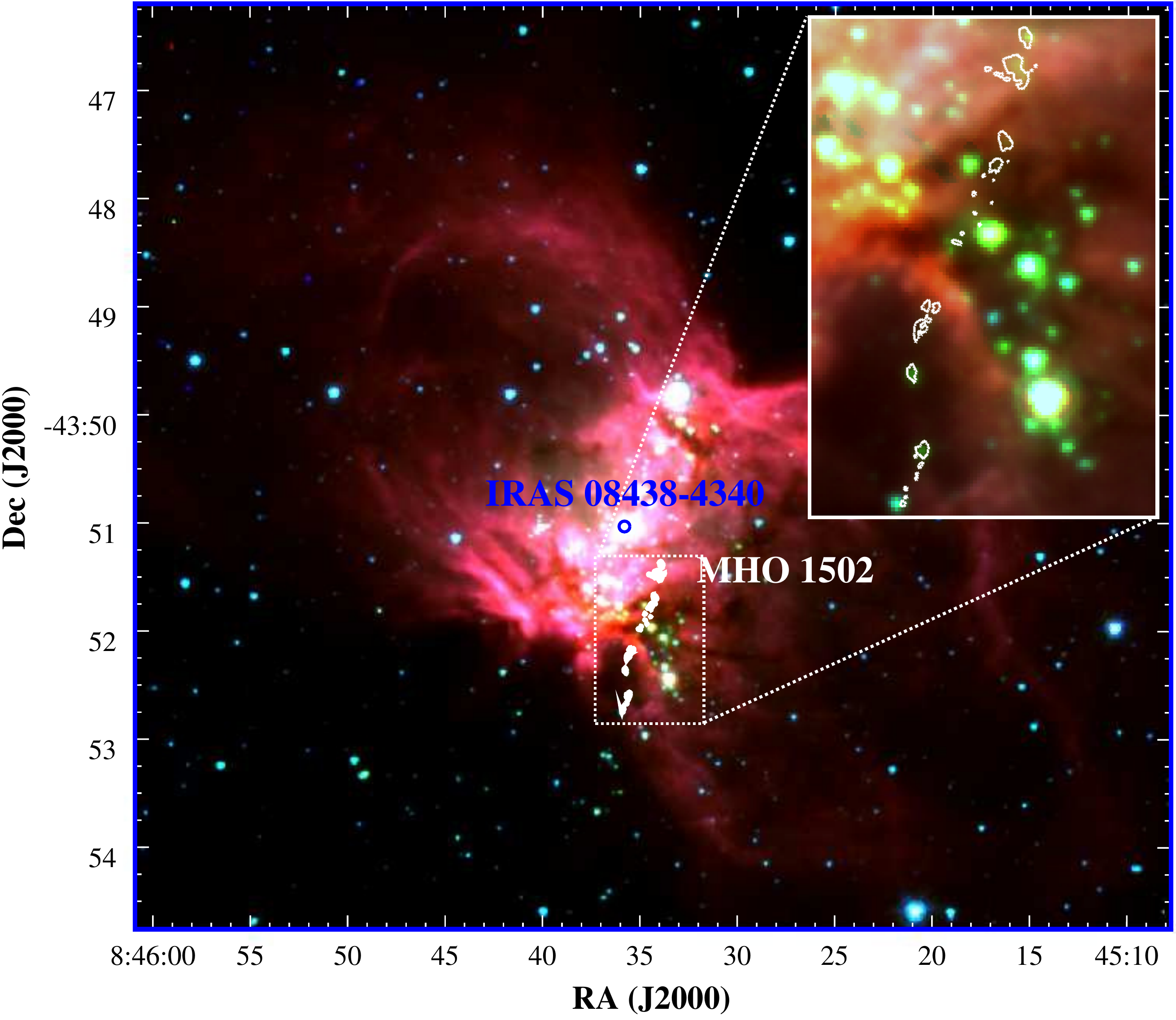}
  \includegraphics[width=\columnwidth]{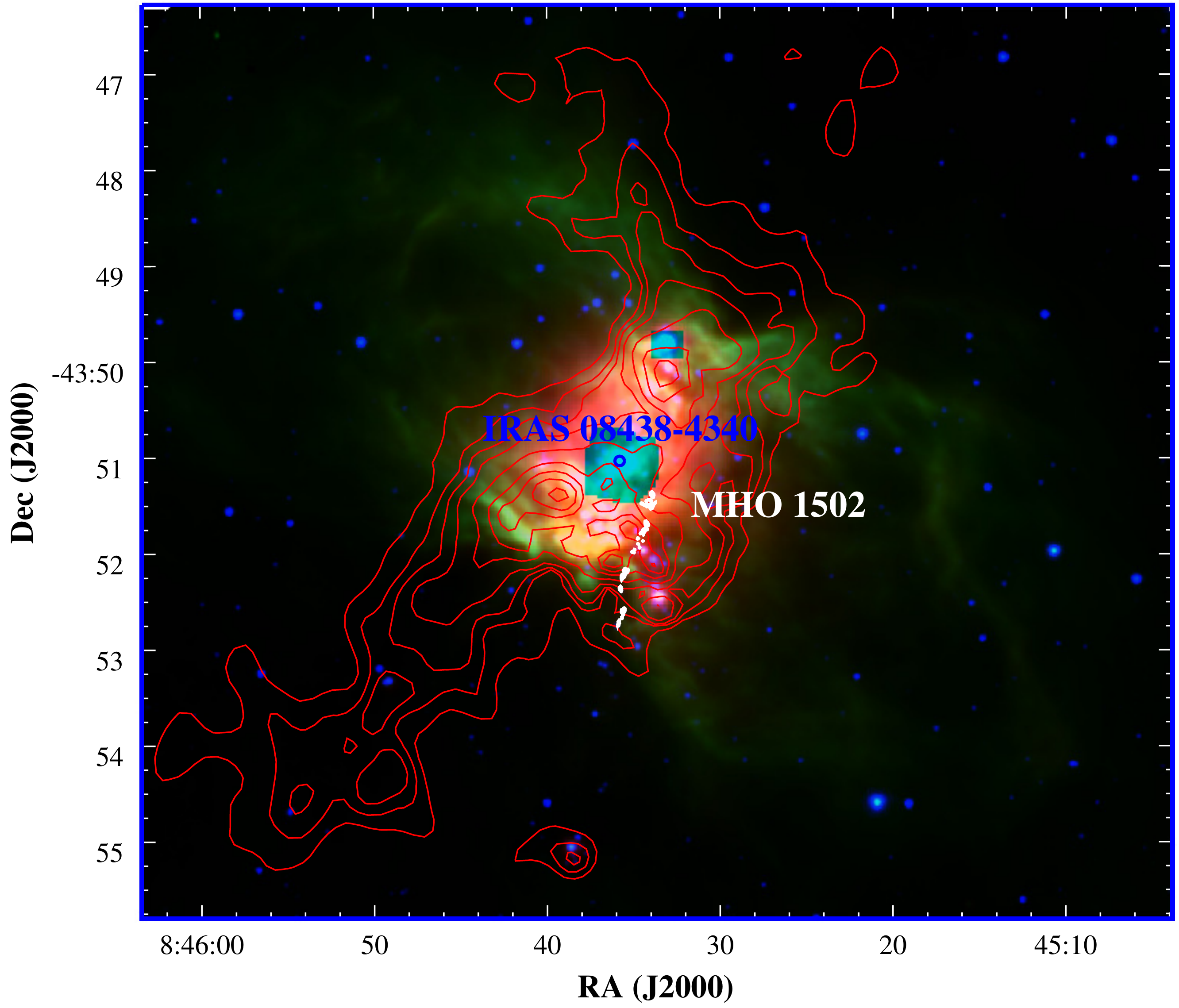}
  \caption{Composite images of the \HII\ region G263.619$-$0.53. \textit{Left panel:} Three--colour image combining IRAC/\Spitzer\ 3.6~\mum\ (blue), 4.5~\mum\ (green) and 8.0~\mum\ (red) images. The inserted enlarged rectangle in the upper-right corner shows in white contours the location of the \H2 knots associated with the MHO~1502 jet. \textit{Right panel:} Combined IRAC-MIPS/\Spitzer\ 4.5~\mum\ (in blue), 8~\mum\ (in green) and 24~\mum\ (in red) images. The superimposed continuous red lines are the column density contours taken from \cite{Marsh2017}. Column density levels are 5, 6, 7, 9, 15, 25, 35, 45 and $56\times10^{21}$~cm$^{-2}$. The white contours indicate the location of the MHO~1502 jet. In both planes, the blue circle marks the position of the IRAS~08438$-$4340 exciting source.}
  \label{fig_mho1502_hii_spitzer}
 \end{figure*}

The knotted structure of several jets, in particular parsec-scale jets, has been interpreted as a `fossil record' of outbursts or FU Orionis events \citep{Hartmann1996} that the exciting sources  have undergone in the past \citep{Reipurth1985,Reipurth1989,Reipurth1997,Reipurth-Bally2001,McGroartyRay2004}. However, the elapse of time between ejections or  knots ($\sim10^3$~yr) is about an order of magnitude less than that of the FU Orionis outbursts ($\sim10^4$~yr), which casts doubt on the FU Orionis status of the knotty-jet exciting sources \citep{Ioannidis-Froebrich2012,Reiter2015,Froebrich2016,Reiter2017}. The knotted structure of the MHO~1502 jet profile, with pairs of quasi-opposite emissions, suggests that the ejections from the exciting source have been intermittent. However, there is currently no evidence to support the FU Orionis status of the proposed exciting source IRAC~18064. In addition, considering the range of distances between consecutive knots (3.7\arcsec\ for knots I and J, and 16\arcsec\ for knots E and F, see Fig.~\ref{fig_MHO1502}), and assuming an average jet velocity of 50-100~\kms\ \citep[see, e.g.][]{Bally2016}, then the elapse of time between knots ranges from $\sim120$ to 1100~yr, which is rather short in comparison with FU~Ori outbursts, but in agreement with estimations obtained for other jets (see references above). However, we caution that the position of IRAC~18064 does not coincide with the centre of the jet, and also that its identification as the driving star is uncertain, as we discuss below in Sect.~\ref{sec_models_masciadri}.

\subsubsection{\texorpdfstring{\H2}{} emissions not associated with MHO~1502: the nearby \texorpdfstring{\HII}{} region \texorpdfstring{G263.619$-$0.53}{} and the IRS 16 cluster}
\label{sec_adj1502}

Figure~\ref{fig_MHO1502} shows numerous (faint) \H2 emissions all over the field (see the yellow arrows), but these are unlikely to be associated with the MHO~1502 jet. Diffuse \H2 emission in the vicinity of massive-star forming regions can be produced by two different mechanisms: excitation by Lyman and Werner UV photons or through collisional excitation due to shocks arising from outflows from nearby young (massive) stars. 

Lyman and Werner UV photons from massive stars penetrate into the region of neutral gas where \H2 exists. Then, UV photons with energies greater than approximately 10~eV can excite \H2 from the ground electronic state to the Lyman and Werner bands. When the \H2 decays down to the ground state, it can emit quadrupole radiation ($\Delta$J~$=-2$) to the J~$=1$ $\nu = 0$ level, giving the line at 2.12~\mum, which traces the interface where UV radiation encounters the molecular hydrogen and delineates the photodissociation region (PDR). Another good tracer of PDRs is the Spitzer 8~\mum\ band \citep[see e.g.][]{Berne2009,Stock2016}, because there are several polycyclic aromatic hydrocarbons (PAHs) in this band. The PAHs are destroyed in the ionised gas, but are thought to be excited in the PDR by the absorption of far-UV photons (6~eV~$<$~h$\nu$~$<$~13.6~eV) at the interface of the \HII\ region and the molecular cloud \citep{Giard1994}.

The molecular hydrogen line at 2.12~\mum\ is also a well-known tracer of a shock region produced by the outflows that sweep the surrounding cloud material \citep{Wakelam2017}. These outflows are ejected by young stars at the very early stage of evolution, with the shock regions usually aligning along the rotation axis and being considered signposts of star formation.
The dispersed \H2\ emissions in the field of the MHO~1502 jet (see Fig.~\ref{fig_MHO1502}) are very close to the \HII\ region G263.619$-$0.53 and the young stellar cluster IRS~16 \citep{Caswell-Haynes1987,Massi2003} centred near to the position of the IRAS~08438$-$4340 source and located $\sim54$\arcsec\ or 0.2~pc (at a distance of 700~pc) to the NE of the centre of this jet.

The left and right panels of Figure~\ref{fig_mho1502_hii_spitzer} display composite images of the \HII\ region \hbox{G263.619$-$0.533}. The white contours mark the positions of the \H2 knots delineating the MHO~1502 jet, revealing the close vicinity between the jet and the \HII\ region. In particular, the right panel of Fig.~\ref{fig_mho1502_hii_spitzer} distinctly shows the bipolar morphology of this \HII\ region to have two lobes that shine at 8~\mum. Moreover, the central region of the bipolar nebula is very bright and saturates at 24~\mum, while the waist is visible as a dark elongated patch and more clearly depicted by the \H2 density contour maps (in continuous red line) of \cite{Marsh2017}. These features are typical of bipolar \HII\ regions, as identified by \citet{Deharveng2015} and \citet{Samal2018}, with the \HII\ region ionising source being an early B0--B2 V star \citep{Massi2003}. According to \cite{Massi2010}, this \HII\ region has not reached its Str\"{o}mgren radius. These authors estimated a dynamical time of $\sim4\times10^{4}$~yr, which is consistent with the age of the infrared cluster IRS~16 (2~Myr) associated with it.

 \begin{figure*}        
  \centering
  \includegraphics[width=\columnwidth]{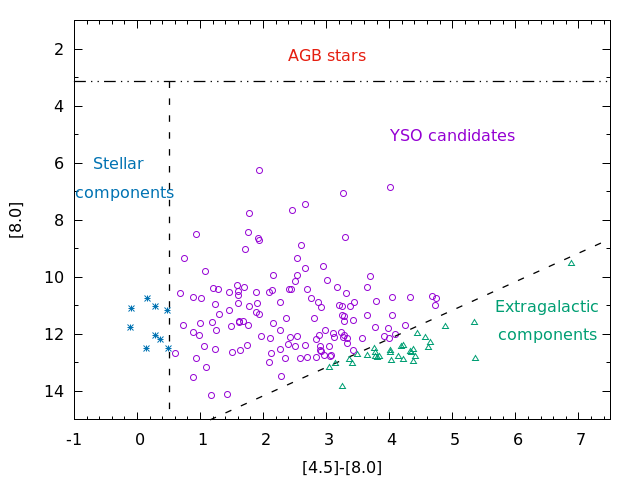}
  \includegraphics[width=\columnwidth, trim={0cm 0cm 1cm 0cm}, clip]{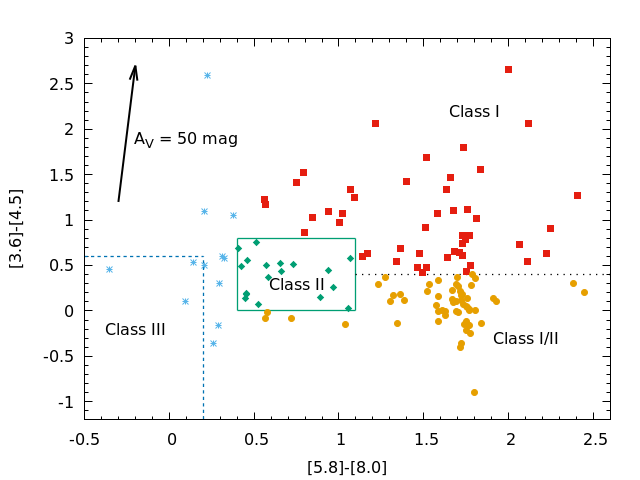}
  \caption{\textit{Left panel:} [8.0] vs [4.5]$-$[8.0] diagram of a subsample of the VMR-D IRAC point-source, catalogued in Table~5 of \cite{Strafella2010}, within a radius of 1.5\arcmin\ centred at the position of MHO~1502 \hbox{($\alpha$,$\delta$)(J2000)\,=\,(08:45:34.2; $-$43:51:54.1)}. The dashed lines mark the statistical criteria of \cite{Harvey2006,Harvey2007} scaled to the distance of VMR-D (see text) in order to distinguish YSO candidates (magenta circles) from other types of contaminating sources, such as: extragalactic objects (green triangles), normal stars (blue asterisks), and AGB stars. 
  \textit{Right panel:} [5.8]$-$[8.0] vs. [3.6]$-$[4.5] diagram. The central squared green region indicates the domain of Class~II YSO sources (green diamonds) according to statistical criteria derived by \cite{Allen2004} and \cite{Megeath2004}. The dotted blue line encloses the location of sources without IR excess \citep{Flaherty2007}. The Class~I (red squares), Class~II (green diamonds), Class~I/II (orange circles), and Class~III/foreground/background sources (light blue asterisks) objects are marked. The reddening vector (black arrow) corresponds to A$_V=50$\,mag \citep{Strafella2010}.}
  \label{fig_color-color}
 \end{figure*}

\begin{figure*}
 \centering
 \includegraphics[width=\textwidth]{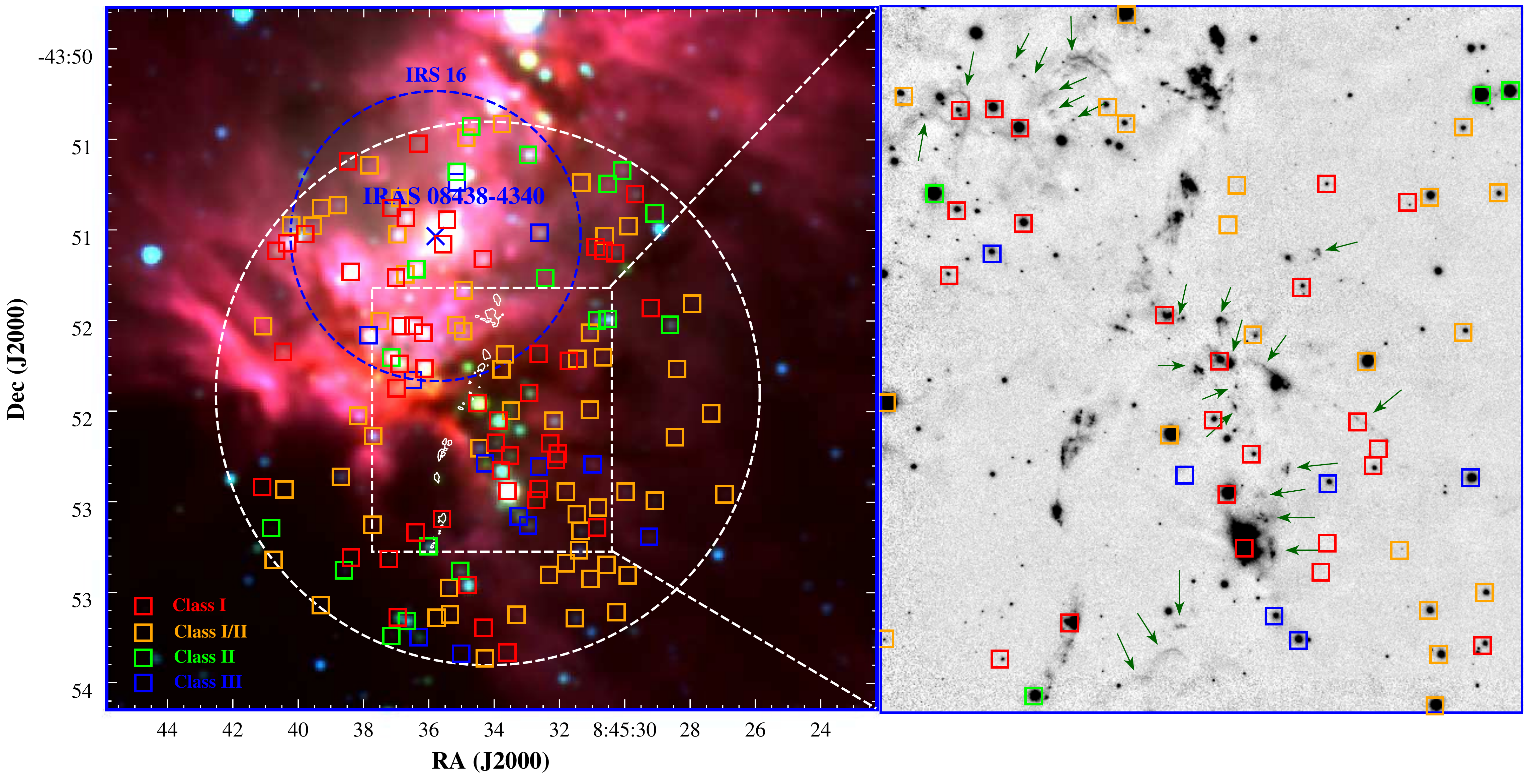}
 \caption{\textit{Left panel:} Three-colour image with 3.6~\mum\ (blue), 4.5~\mum\ (green) and 8.0~\mum\ (red) of the \HII\ region G263.619$-$0.53.  The dashed blue circle indicates the area of the IRS~16 stellar cluster defined by \cite{Massi2003}.  The dashed white square highlights the area observed with GSAOI+GeMS, comprising the MHO~1502 jet.  The dashed white circle is centred at \hbox{($\alpha$,$\delta$)(J2000)\,=\,(08:45:34.2; $-$43:51:54.1)}, has a radius of 1.5\arcmin, and defines our search area. The white contours mark the \H2 knots of MHO~1502 and the blue cross indicates the location of \hbox{IRAS~08438$-$4340}, the central source associated with the \HII\ region. The coloured squares show the YSOs detected in the [8.0] vs [4.5]$-$[8.0] diagram and are classified according to the [5.8]$-$[8.0] vs [3.6]$-$[4.5] diagram (see Fig.~\ref{fig_color-color}) as follows: Red squares are Class~I, orange squares denote Class~I/II, and green ones are Class~II sources. Blue squares correspond to Class~III as well as the foreground and background sources.
 \textit{Right panel:} Inverted grey-scale \H2 image of the MHO~1502 jet obtained with GSAOI$+$GeMS. YSOs in the field are shown. The symbols used are the same as those in the left panel. The dark green arrows highlight the \H2 emissions not associated with the MHO~1502 jet (see also Fig.~\ref{fig_MHO1502}).}
 \label{fig_mho1502_ysos}
\end{figure*}

Several bipolar \HII\ regions are associated with the star formation taking place in the flat or sheet-like cold molecular material delineating the waist of the nebula \citep{Dewangan2016,Dewangan2019}. \cite{FukudaHanawa2000} suggested that the expansion of an \HII\ region near a filamentary molecular cloud can generate sequential waves of star formation. Indeed, the star formation process can be triggered by the compression of the molecular material swept up by the ionisation front or the squeezing of a pre-existing dense cloud. The existence of the infrared cluster IRS 16 (indicated with a dashed blue circle in Fig.~\ref{fig_mho1502_ysos}), associated with this \HII\ region lends support to this star-formation scenario. The projection of the MHO~1502 jet onto the SW lobe of the bipolar \HII\ region is clearly seen in Fig.~\ref{fig_mho1502_hii_spitzer}, again suggesting an ongoing star-formation process in the region.  

Based on the spatial distribution of the \H2 emissions in the field of Fig.~\ref{fig_MHO1502} and the location of the nearby \HII\ region G263.619$-$0.533 and the infrared cluster IRS~16, we tried to disentangle the origin of the \H2 emissions in the field that are not associated with the MHO~1502 jet. To investigate the young stellar objects (YSOs) in the region as candidate driving sources of the adjacent emissions in Fig.~\ref{fig_MHO1502}, we used the \Spitzer-IRAC/MIPS photometric point-source catalogue of the Vela cloud complex D (VMR-D), published by \cite{Strafella2010}. From their Table~5, we selected a subsample within a 1.5\arcmin\ radius centred at the position of MHO~1502 \hbox{($\alpha$,$\delta$)(J2000)\,=\,(08:45:34.2; $-$43:51:54.1)}. In the left panel of Fig.~\ref{fig_mho1502_ysos}, the dashed white circle indicates our search area (within a radius of 1.5\arcmin), the dashed white square shows the area of the IRS~16 cluster, enclosed within a radius of $\sim 0.73\arcmin$ or 0.15~pc at a distance of 700~pc \citep{Massi2003}. The white contours mark the location of the \H2 knots of MHO~1502, and the proximity between the jet and the cluster is clear.

Using the $[8.0]$~versus~$[4.5]\,-\,[8.0]$ colour--magnitude diagram (CMD; see Fig.~\ref{fig_color-color}, left panel) and the statistical criteria of \cite{Harvey2006,Harvey2007} scaled to the distance of VMR-D, we identified YSO candidates ($[4.5]\,-\,[8.0]\,>\,0.5$ and $[8.0]\,<\,16.15\,-\,[4.5]\,-\,[8.0]$) from other sources, such as: stellar components ($[4.5]\,-\,[8.0]\,<\,0.5$), extragalactic objects ($[8.0]\,>\,16.15\,-\,[4.5]\,-\,[8.0]$), and asymptotic giant branch (AGB) stars ($-0.86\,<\,[8.0]\,<\,3.13$). We then used the $[3.6]\,-\,[4.5]$ versus $[5.8]\,-\,[8.0]$ diagram (see Fig.~\ref{fig_color-color}, right panel) to place the YSO candidates identified in the CMD, and applied the criteria of \cite{Allen2004} and \cite{Megeath2004} to classify them according to three different proto- and pre-stellar classes. Class~II candidates (green diamonds) are located within the green box-shaped region. The boundary between Class~I (red squares) and Class~I/II (orange diamonds) is indicated by a continuous dotted black line. The region of Class~III candidates, and also likely stellar sources without infrared excesses (light blue asterisks) is shown with a dashed blue line \citep{Flaherty2007}. The black arrow indicates the reddening vector corresponding to $A_V=50$~mag \citep{Strafella2010}.

In Fig.~\ref{fig_mho1502_ysos} the YSO candidates are superimposed with different coloured squares according to their classification: Class~I (red), Class~I/II (orange), Class~II (green), and Class~III\,(\,foreground and\,background) (blue), with many YSO candidates being spread over our search area. The right panel is the inverted grey-scale \H2 image observed with GSAOI$+$GeMS (see Fig.~\ref{fig_MHO1502}). The \H2 emissions in the middle and lower part of the field, marked with dark green arrows, lie close to YSOs in the field. Their proximity to the YSOs suggests that these are shock-excited regions associated with outflows from the YSOs in the region. On the other hand, in the upper left corner, there are several arc- or bow-shock-shaped \H2 structures showing the brightest apexes roughly facing towards the centre of the bipolar \HII\ region, where the ionising source is located. These arc-shape structures in \H2 do not seem to align or proceed from any Class I or I/II sources in the region. This morphology suggests that at least some of these \H2 emissions could have been created by UV fluorescence photons \citep{Deharveng2010,Hartigan2012}, although an outflow origin cannot be ruled out. The three arc-shape emissions located at the bottom of the field with no nearby YSOs are likely to have been produced by UV fluorescence photons.

\begin{figure*} 
    \centering
    \includegraphics[width=\textwidth]{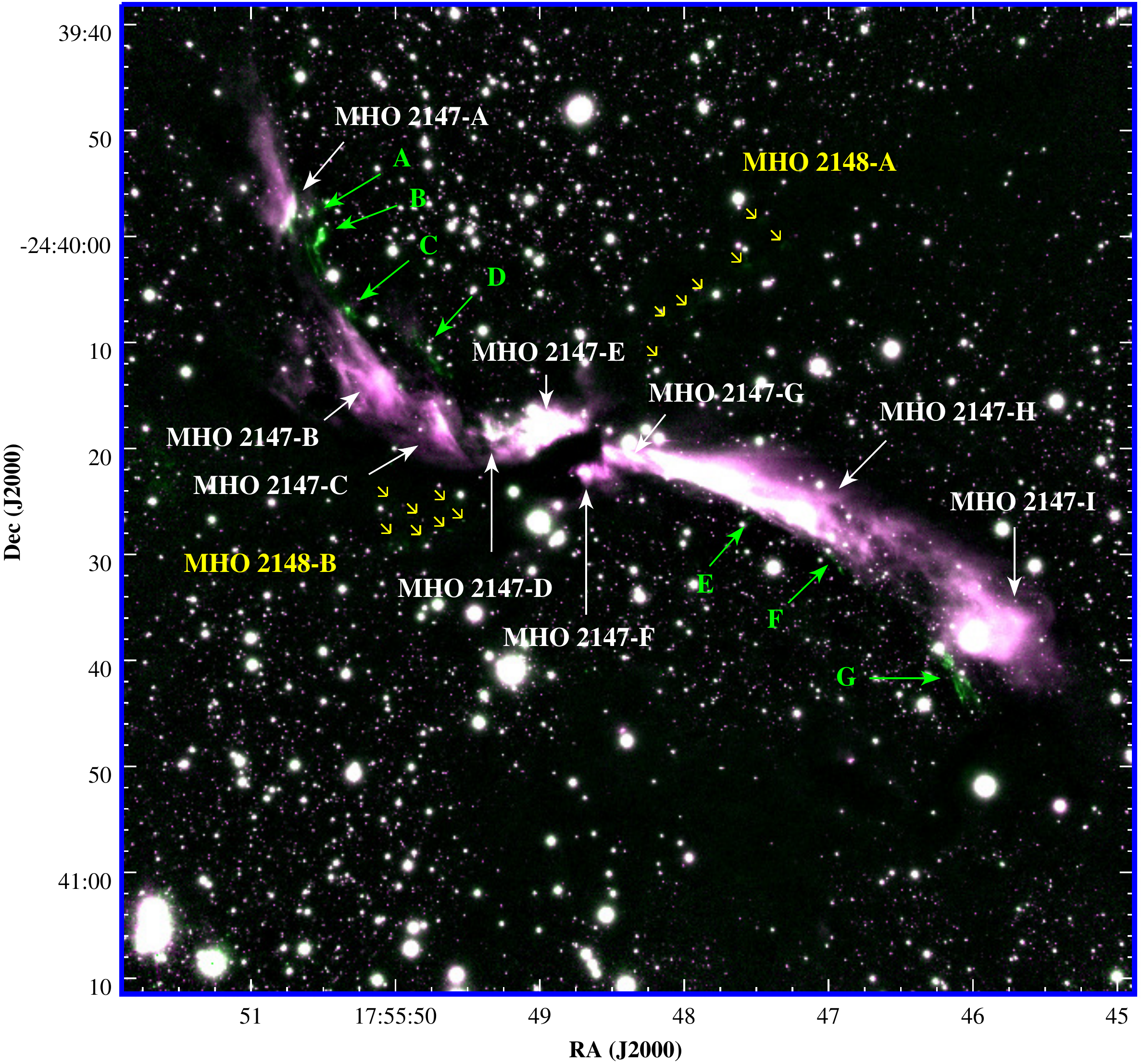}
    \caption{Composite image of MHO~2147 obtained with GSAOI/GEMINI. The K-band filter is in magenta and the \H2-band filter is in green. White arrows mark the position of the different knots associated with MHO~2147. Green arrows highlight the knots that seem to belong to another jet (designated Ad--jet) lying adjacent to MHO~2147, while yellow arrows indicate the location of fainter knots linked to the quasi-perpendicular jet (with respect to MHO~2147) MHO~2148. Both the Ad--jet and MHO~2148 were previously reported by \cite{Varricatt2011}.}
    \label{fig_MHO2147}
\end{figure*}
\begin{table*}
\centering
\begin{threeparttable}
 \caption{Coordinates and \texorpdfstring{\H2}{} fluxes for the knots associated with the MHO~2147 jets.}
 \label{tab_photo2147} 
 \sisetup{table-number-alignment = center, table-text-alignment = center}
 \begin{tabular}{llccD{+}{\,\pm\,}{-1}cl}
 \hline 
 \hline \noalign{\smallskip}
 \multicolumn{2}{c}{\multirow{2}{4em}{Knot ID}}  &
  $\alpha$ (J2000.0)    &
  $\delta$ (J2000.0)    &  
  \multicolumn{1}{c}{Flux}                  &
  \multicolumn{1}{c}{r\tablefootmark{a}}                   & 
  \multirow{2}{10em}{Fig. Reference\tablefootmark{b}} \\
             &                  & 
    ($^{h}:^{m}:^{s}$)      & 
    (\degr:\arcmin:\arcsec) &  
    \multicolumn{1}{c}{($10^{-6}$ Jy)} & 
    \multicolumn{1}{c}{(\arcsec)}  &  \\
  \hline
\noalign{\smallskip}
            A & \,  & 17:55:50.76 & -24:39:57.7 & 170.39+34.52 & 2.12 &  \ref{fig_MHO2147_ABCDEFG}, upper left panel \\
          \hline
\noalign{\smallskip}
            B & 1 & 17:55:50.11 & -24:40:13.4 & 107.77+9.91  & 1.20 & \multirow{2}{10em}{\ref{fig_MHO2147_ABCDEFG}, upper right panel}  \\
            & 2 & 17:55:49.97 & -24:40:14.3 & 5.61+0.49    & 0.30 &   \\
          \hline
\noalign{\smallskip}
           C & \,  & 17:55:49.69 & -24:40:17.5 & 93.62+7.47   & 1.18 &  \ref{fig_MHO2147_ABCDEFG}, middle left panel \\
          \hline
\noalign{\smallskip}
          D & 1 & 17:55:49.36 & -24:40:18.3 & 13.17+1.00   & 0.30 &  \multirow{3}{10em}{\ref{fig_MHO2147_ABCDEFG}, middle right panel} \\
            & 2 & 17:55:49.34 & -24:40:19.0 & 17.23+1.30   & 0.33 &   \\
            & 3 & 17:55:49.28 & -24:40:18.8 & 15.46+1.25   & 0.39 &   \\
          \hline
\noalign{\smallskip}
          E & 1 & 17:55:49.10 & -24:40:19.1 & 46.50+9.34   & 0.60 &  \multirow{3}{10em}{\ref{fig_MHO2147_ABCDEFG}, lower left panel} \\
            & 2 & 17:55:48.98 & -24:40:17.8 & 74.24+5.32   & 0.38 &   \\
            & 3 & 17:55:48.97 & -24:40:18.3 & 71.74+5.13   & 0.33 &   \\
          \hline
\noalign{\smallskip}
          F & \,  & 17:55:48.70 & -24:40:22.3 & 54.89+11.47  & 1.05 &  \ref{fig_MHO2147_ABCDEFG}, lower right panel \\
          \hline
\noalign{\smallskip}
          G & 1 & 17:55:48.53 & -24:40:20.6 & 43.87+13.58   & 0.75 & \multirow{2}{10em}{\ref{fig_MHO2147_ABCDEFG}, lower right panel} \\
          
            & 2 & 17:55:48.37 & -24:40:20.7 & 99.13+20.42  & 1.27 &   \\
          \hline
\noalign{\smallskip}
          H & \,  & 17:55:47.97 & -24:40:22.4 & 2005.87+183.03 & 5.38 &  \ref{fig_MHO2147_HI}, left panel \\
          \hline
\noalign{\smallskip}
           I\tablefootmark{c} &    & 17:55:45.78 & -24:40:37.4 & \multicolumn{1}{c}{$-$} & {--}& \ref{fig_MHO2147_HI}, right panel \\
  \hline  
 \end{tabular}  
\tablefoot{
\tablefoottext{a}{Radii of the circular apertures used for photometry.}
\tablefoottext{b}{Reference to the figure and panel in which an enlarged version of each knot can be found.}
\tablefoottext{c}{Flux contaminated by nearby sources.}}
\end{threeparttable}
\end{table*}

\subsection{MHO~2147} 

\subsubsection{Characteristics of the jet}
\label{sec_MHO2147}

The jet MHO 2147, located in the Ophiuchus region at a distance of 3.23~kpc \citep{Molinari1996}, was discovered in \H2 by \cite{Varricatt2011}. The luminous Class~I source IRAS~17527$-$2439 (hereafter IRAS~17527) is the proposed exciting star of the jet. \cite{Chen2013a} associated IRAS~17527 with the {\it Extended Green Object} \object{EGO~G4.83$+$0.23} \citep{Cyganowski2008}, and reported the detection of this EGO in 95~GHz, a Class~I methanol maser and a known tracer of shock gas associated with outflows \citep[e.g.][]{Cyganowski2009}.

Figure~\ref{fig_MHO2147} shows a combined \H2 (in green) and K (in magenta) image of MHO~2147 obtained with GSAOI$+$GeMS/Gemini. The MHO~2147 jet displays a well-defined, gently wiggling profile extending 1.46~pc from the northeast to the southwest. The projection of the jet on the plane of the sky crosses a dark patch, which is clearly visible in the middle of the figure and has a position angle of 37\degr. Knots associated with MHO~2147 are marked with white arrows and labelled with capital letters. A significant amount of emission in the K-band, which coincides with the \H2 emission, can be seen along the jet, particularly in the central region where the knots E, F, and G and the southwest lobe are located. \cite{Varricatt2011} observed the same effect in JHK and IRAC/\Spitzer\ images. This characteristic is typical of massive young stellar objects (MYSOs) surrounded by a massive accretion disc, and is probably due to the reflection of scattered light from the central source into outflow cavity walls \citep[e.g.][]{Reipurth2000,Arce2007}. 

In addition, Figure~\ref{fig_MHO2147} shows several knots (indicated with green arrows) that were only detected in the \H2\ filter, with no counterpart appearing in the K-band, which appear in green in this figure. These are slightly offset from the main path of the jet, but otherwise follow the extension and profile of the S-shape MHO~2147 jet. These knots seem to belong to a jet adjacent to MHO 2147, hereafter referred to as Ad--jet. The NE lobe of the Ad--jet is shifted by $\sim15\degr$, while the SW lobe is displaced by $\sim8\degr$ with respect to MHO~2147.

The knots identified as belonging to MHO~2147 are bright and extended, whereas those of the Ad--jet are much fainter and point-like, and in some cases are surrounded by a small nebulous emission. As mentioned above, the knots in the MHO~2147 jet spatially coincide with K-band emission, while those in the Ad--jet are of purely \H2\ emission. On average, the distance between consecutive knots in the MHO~2147 jet is much larger than for the Ad--jet. It is unlikely that these differences in knot brightness, morphology, and spacing are produced by a unique exciting source, and suggest that each jet is probably driven by different exciting sources. 

Enlarged \H2 flux calibrated images of the NE and SW knots along MHO~2147 and the Ad--jet can be found in Figs.~\ref{fig_MHO2147_ABCDEFG}, \ref{fig_MHO2147_HI}, and \ref{fig_ad-jet_BDG}. The substructures in each knot are marked and labelled with Arabic numbers, as mentioned in Sect.~\ref{sec_observations}. The knot morphologies are relatively diverse, with some, such as in the cases of the knots MHO~2147-A and Ad--jet~B (Fig.~\ref{fig_MHO2147_ABCDEFG}, upper left panels, and \ref{fig_ad-jet_BDG}, left panel), showing several localised (point-like) emissions grouped in a small area and surrounded by a more diffuse and fainter emission. In other cases, the \H2 diffuse emission delineates the knot (elongated) structure, with only one central \H2 condensation, as in the cases of knots \hbox{MHO~2147-B} and C (Fig.~\ref{fig_MHO2147_ABCDEFG}, upper right and middle left panels). Knot \hbox{MHO~2147-H} presents a very elongated shape of about 0.35~pc in length ($\sim22.5$\arcsec, see Fig.~\ref{fig_MHO2147_HI}, left panel) whereas knot G of the Ad--jet exhibits a peculiar shape composed of several filaments (with a PA of 213.5\degr, see Fig.~\ref{fig_MHO2147}). The enlarged image of this latter knot (Fig.~\ref{fig_ad-jet_BDG}, bottom panel) reveals several individual (point-like) emissions in \H2. Tables~\ref{tab_photo2147} and \ref{tab_photoAd_jet} list identifications of the MHO~2147 and Ad--jet knots.

In addition, \cite{Varricatt2011} detected a much fainter jet, MHO~2148 \citep{Davis2010}, which lies almost perpendicular to MHO~2147 in the plane of the sky, and suggested that MHO~2148 could be related to a companion source of IRAS~17527. Figure~\ref{fig_MHO2147} shows a chain of faint knots highlighted with yellow arrows, delineating the MHO~2148 jet. This jet seems to come from the same origin as MHO~2147. Figure~\ref{fig_MHO2148_AB}, left and right panels, shows an enlarged composite image (K-band filter in magenta and \H2-band filter in green) of the MHO~2148 jet. The north-west lobe (MHO~2148-A) extends $\sim0.47$~pc (30\arcsec) with a P.A of 313\degr, while the southeast lobe (MHO~2148-B) stretches for $\sim0.3$~pc (19\arcsec), with a P.A of 122\degr. Table~\ref{tab_mho2148} lists the coordinates of the detected \H2 knots.

\begin{table*}
\centering
\begin{threeparttable}
 \caption{Coordinates and \texorpdfstring{\H2}{} fluxes for the knots associated with the Ad--jet.}
 \label{tab_photoAd_jet} 
 \sisetup{table-number-alignment = center, table-text-alignment = center}
 \begin{tabular}{
        llcc
        S[separate-uncertainty = true,
          table-figures-uncertainty = 1,
          table-figures-decimal = 2,
      table-figures-integer = 2]
        S[table-format = 1.2]  
        l}
 \hline 
 \hline \noalign{\smallskip}
  \multicolumn{2}{c}{\multirow{2}{4em}{Knot ID}}  & $\alpha$ (J2000.0) & $\delta$ (J2000.0)      &  Flux           &  r\tablefootmark{a} & \multirow{2}{10em}{Fig. Reference\tablefootmark{b}} \\
             &                  & ($^{h}:^{m}:^{s}$) & (\degr:\arcmin:\arcsec) &  {($10^{-6}$ Jy)} & {(\arcsec)}  &  \\
  \hline
\noalign{\smallskip}
          A &   & 17:55:50.58 & -24:39:57.8 & 13.90\pm 2.80     & 0.54 & \ref{fig_MHO2147_ABCDEFG}, upper left panel \\
          \hline
\noalign{\smallskip}
          B & 1 & 17:55:50.52 & -24:39:59.4 & 15.09\pm 1.11   & 0.35 &  \multirow{3}{10em}{\ref{fig_ad-jet_BDG}, upper left panel} \\
            & 2 & 17:55:50.52 & -24:40:00.2 & 21.16\pm 1.53   & 0.35 &   \\
            & 3 & 17:55:50.55 & -24:40:00.7 & 6.18\pm 0.46    & 0.23 &   \\
          \hline
\noalign{\smallskip}
          C &   & 17:55:50.34 & -24:40:06.8 & 10.94\pm 0.93   & 0.40 & \ref{fig_MHO2147_ABCDEFG}, upper right panel \\  
          \hline
\noalign{\smallskip}
          D & 1 & 17:55:49.85 & -24:40:10.2 & 0.87\pm 0.07    & 0.15 &  \multirow{3}{10em}{\ref{fig_ad-jet_BDG}, right panel}\\
            & 2 & 17:55:49.76 & -24:40:11.7 & 1.22\pm 0.25    & 0.21 &   \\
            & 3 & 17:55:49.71 & -24:40:12.6 & 1.68\pm 0.14    & 0.22 &   \\
          \hline
\noalign{\smallskip}
          E\tablefootmark{c} &   & 17:55:47.56 & -24:40:26.6 & {--}  & {--} &  \ref{fig_MHO2147_HI}, left panel \\
          \hline
\noalign{\smallskip}
          F & 1\tablefootmark{c} & 17:55:46.98 & -24:40:30.3 & {--} & {--} &  \multirow{2}{10em}{\ref{fig_MHO2147_HI}, left panel} \\
            & 2 & 17:55:46.93 & -24:40:31.4 & 3.29\pm 0.10 & 0.40 &   \\
          \hline
\noalign{\smallskip}
          G & 1  & 17:55:46.18 & -24:40:39.6 & 2.41\pm 0.21   & 0.18 &  \multirow{8}{10em}{\ref{fig_ad-jet_BDG}, bottom panel}  \\
            & 2  & 17:55:46.20 & -24:40:40.0 & 1.43\pm 0.12   & 0.12 &   \\
            & 3  & 17:55:46.18 & -24:40:40.1 & 0.73\pm 0.06   & 0.08 &   \\
            & 4  & 17:55:46.21 & -24:40:40.5 & 1.00\pm 0.08   & 0.12 &   \\
            & 5  & 17:55:46.04 & -24:40:42.1 & 1.26\pm 0.11   & 0.13 &   \\
            & 6  & 17:55:46.10 & -24:40:42.3 & 0.81\pm 0.08   & 0.12 &   \\
            & 7  & 17:55:46.09 & -24:40:42.5 & 0.62\pm 0.07   & 0.11 &   \\
            & 8  & 17:55:46.09 & -24:40:43.5 & 1.07\pm 0.09   & 0.11 &   \\
  \hline  
 \end{tabular}  
\tablefoot{
\tablefoottext{a}{Radii of the circular apertures used for photometry.}
\tablefoottext{b}{Reference to the figure and panel in which an enlarged version of each knot can be found.}
\tablefoottext{c}{Flux lower than 3$\sigma$ rms.}}
\end{threeparttable}
\end{table*}

\begin{figure}
    \centering
    \includegraphics[width=\columnwidth]{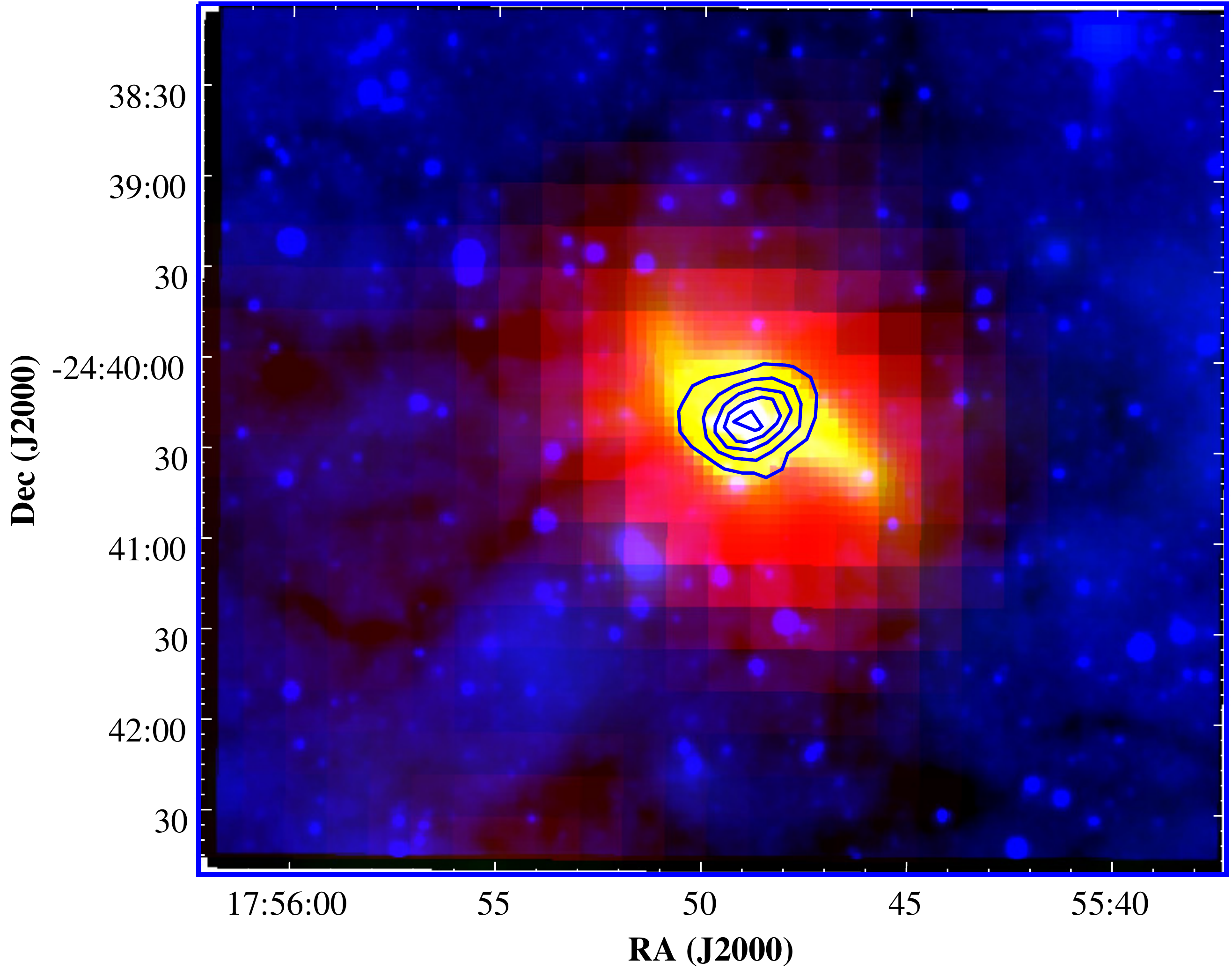}
    \caption{Composite image of a field of $10'\times7'$ centred on IRAS~17527$-$2439. The IRAC/\Spitzer\ 8~\mum\ band is in blue, PACS/\Herschel\ 70~\mum\ is in yellow, and SPIRE/\Herschel\ 500~\mum\ is in red. The blue contours are from the ATLASGAL 870~\mum\ data, and correspond to levels of 0.6, 0.98, 1.35, 1.73, and 2.1~Jy~beam$^{-1}$. The patchy opaque silhouettes extending towards the border of the figure highlight the IRDC detected by \cite{Chen2013a}, associated with the EGO~G4.83$+$0.23.}
    \label{fig_mho2147_atlasgal}
\end{figure}

\subsubsection{The mass of the dark cloud}
\label{sec_mass-clump_mho2147}

IRAS~17527 is located 2.8\arcmin\ NW from the position of a dark cloud (\object{Dobashi~0238}) catalogued by \cite{Dobashi2011}. \cite{Chen2013a} associated EGO~G4.83$+$0.23, and thus also IRAS~17527, with an infrared dark cloud (IRDC\footnote{The infrared dark clouds (IRDCs) are usually identified as extended opaque silhouettes seen in contrast to the bright mid-IR Galactic background emission \citep{Egan1998}. In particular, these structures are detected as dark clouds at 8~\mum\ \citep{Simon2006a,Peretto2009}. They are cold (T~$<20$~K), have high molecular hydrogen column densities (N(\H2)~$\sim 10^{22}$~cm$^{-2}$), large masses ($\sim 10^2-10^5$~\msun), and are sites of high star-formation \citep{Simon2006b,Rathborne2006,Pillai2006}.}). \cite{Contreras2013} and \cite{Urquhart2014} identified a source at 870~\mum, catalogued as \object{AGAL~004.827$+$00.231} in the ATLASGAL survey\footnote{\url{http://atlasgal.mpifr-bonn.mpg.de/cgi-bin/ATLASGAL_DATABASE.cgi}} and coinciding with the position of IRAS~17527, indicating the presence of cold dust. Figure~\ref{fig_mho2147_atlasgal} shows a composite image of a field of $10'\times7'$ centred on IRAS~17527. 
The emission at 8~\mum\ is mainly localised in the central region, where IRAS~17527 is found. The patchy opaque silhouettes extending towards the border of the figure highlight the IRDC detected by \cite{Chen2013a}. On the other hand, the emissions at 70~\mum\ and 160~\mum\ extend from the northeast to the southwest, following the trace of the \H2 emission in Fig.~\ref{fig_MHO2147}. At 500~\mum, the emitting region has a size of $\sim3$~pc whereas at 870~\mum\ the contour at 0.6~Jy~beam$^{-1}$ (over $3\sigma$ rms) has a linear dimension of $\sim0.7$~pc.

\begin{figure}
    \centering
    \includegraphics[width=\columnwidth, trim={0cm 0cm 1.5cm 0cm}, clip]{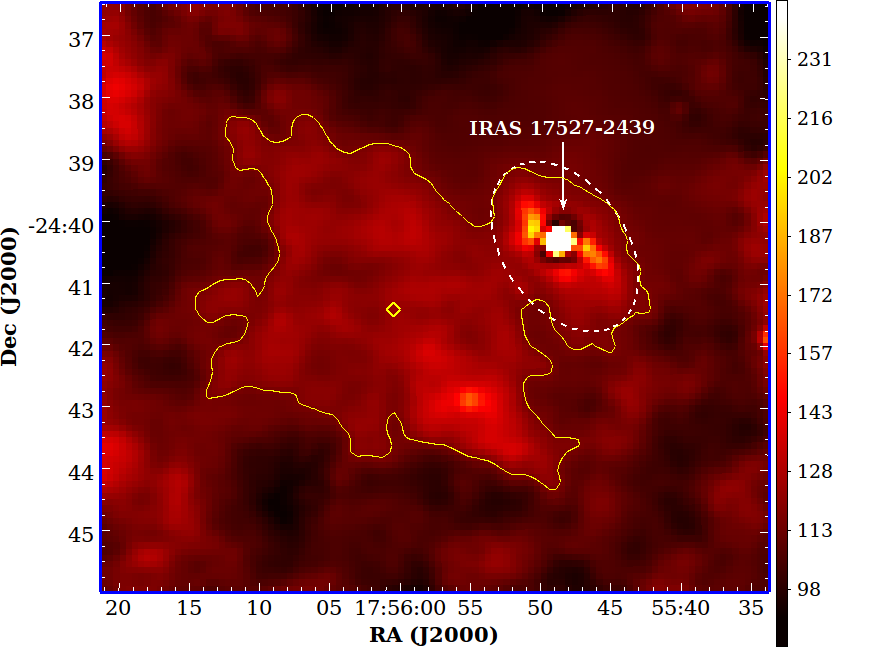}
    \caption{Column density maps in an area of $10.8\arcmin\times 9.5\arcmin$ centred at \radec~$=$~(17:55:57.60, $-$24:41:13.49), obtained from \cite{Marsh2017}. The colour scale is in units of $10^{20}$~cm$^{-2}$. The yellow contour is at $1.155\times10^{22}$~cm$^{-2}$. The dashed white ellipse roughly comprises the IRDC detected by \cite{Chen2013a}, and the yellow diamond indicates the central position of the dark cloud 0238 from \cite{Dobashi2011}.}
    \label{fig_mho2147_dens}
\end{figure}

We used the column density maps from \cite{Marsh2017}\footnote{Available at the web site: \url{http://www.astro.cardiff.ac.uk/research/ViaLactea/}} to estimate the mass of the dark cloud in which IRAS~17527 is embedded. These maps were produced by employing the Bayesian point process mapping (PPMAP) method and applying it to \Herschel\ continuum data in the wavelength range of 70--500\,\mum\ (see \citealt{Marsh2015}). The PPMAP procedure provides resolution-enhanced ($\sim12$\arcsec) image cubes of differential column densities and dust temperatures. Figure~\ref{fig_mho2147_dens} shows the density map N(\H2) in an area of $10.8\arcmin \times 9.5\arcmin$ centred at \hbox{\radec~$=$~(17:55:57.60, $-$24:41:13.49)}. We integrated over the area delimited by the contour of \hbox{$1.155\times10^{22}$}~cm$^{-2}$ at $3\sigma$ (shown with yellow line in Fig.~\ref{fig_mho2147_dens}), which comprises both the Dobashi~0238 dark cloud and the IRDC associated with EGO~G4.83$+$0.23 catalogued by \cite{Chen2013a} with the temperature within this area being found to vary from 20.3 to 26.3~K. We used the expression $M = \mu\, m_{H}\, A\, N_{H_2}$, where $\mu = 2.76$ is the mean molecular weight for a helium abundance of 25\% \citep[per mass,][]{Yamaguchi1999,Miettinen2012}, m$_H$ is the mass of the hydrogen atom ($1.67\times10^{-24}$~g), and $A$ is the area subtended by the yellow line contour in Fig.~\ref{fig_mho2147_dens}. The hydrogen gas mass was calculated to be $M_{H_2} \sim 6300$~\msun. We also estimated the mass enclosed by the dashed white ellipse shown in Fig.~\ref{fig_mho2147_dens}, and this roughly coincided with the IRDC detected by \cite{Chen2013a}. The ellipse is centred at \radec~$=$~(17:55:48.32, $-$24:40:24.15), with an area of $1.5\arcmin \times 1\arcmin$ (inclination of 305\degr), similar to the emission at 500~\mum\ (in red) shown in Fig.~\ref{fig_mho2147_atlasgal}. We obtained a hydrogen gas mass of $\sim 1400$~\msun, which is roughly a factor of five lower than the mass contained by the dark cloud (Dobashi~0238).

\subsubsection{The \texorpdfstring{IRAS~17527$-$2439}{} SED model}
\label{sec_exciting-source_mho2147}

\cite{Varricatt2011} analysed the IRAC colour--colour diagram within an area of $10\arcmin \times 7\arcmin$ centred on the IRAS~17527 source. This author identified approximately 20 Class~II and a handful of Class~I candidate YSOs, including IRAS~17527,  which is a highly reddened Class~I object. In addition, he compiled fluxes from 3.6 to 1200~\mum\ for this IRAS source, and noted a bright (saturated) 24~\mum\ MIPS source at a separation of only 2\arcsec\ from the position of the IRAS source\footnote{Coordinates for the 24~\mum\ MIPS source were derived from the 8~\mum\ IRAC image \citep{Varricatt2011}.}. The author also included, as the upper limit in the K-band, the flux of a deeply embedded object that was only detected in the K and \H2 bands, labelled `source A'. Finally, he constructed and modelled the corresponding SED of IRAS~17527. 

From a literature search, we identified several fluxes at mid-IR, far-IR, and submillimetre wavelengths that were not included in the SED analysed by \citet{Varricatt2011}. In particular, we can mention the 8.28 (A band), 12.13 (C band), and 21.34~\mum\ (E band) fluxes from the Midcourse Space Experiment MSX6C Infrared Point Source Catalog \citep{Egan2003}, 70 and 160~\mum\ data taken from the \Herschel-PACS Point Source Catalogue \citep[HPPSC,][]{Marton2017,Herschel_PACS_Vizier_2020}, the fluxes at 250, 350, and 500~\mum\ obtained from the \Herschel\ Infrared GAlactic Plane Survey \citep[Hi-GAL,][]{Molinari2010}, the flux at 870~\mum\ taken from the APEX Telescope LArge Survey of the GALaxy \citep[ATLASGAL,][]{Schuller2009,Contreras2013,Urquhart2014}, and the flux at 1.1~mm from the Bolocam Galactic Plane Survey \citep[BGPS v2.1,][]{Ginsburg2013}.

\begin{figure}
    \centering
    \includegraphics[width=0.49\textwidth]{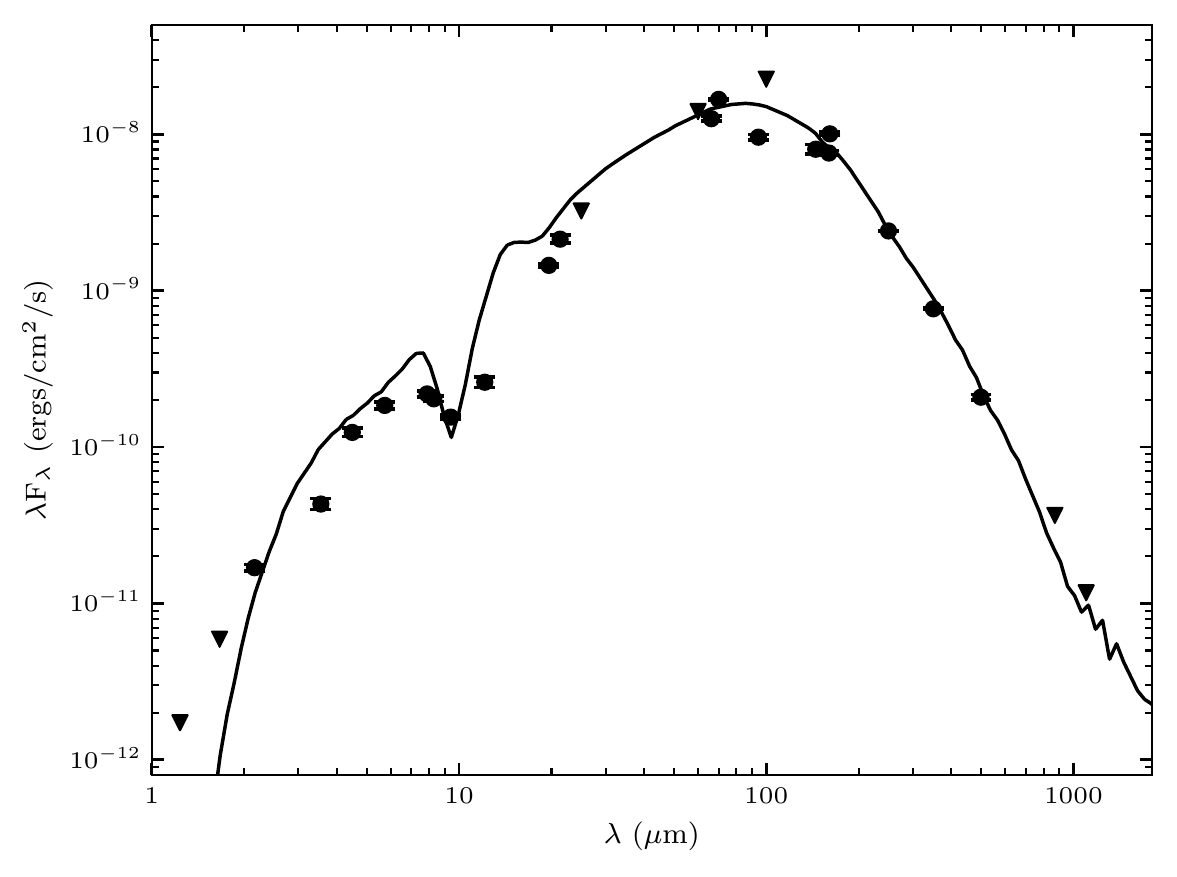}
    \caption{Observed and modelled SED of IRAS 17527$-$2439. The 2MASS fluxes at 1.25~\mum\ (J band) and at 1.66~\mum\ (H band), the IRAS fluxes at 25, 60 and 100~\mum, and the ATLASGAL flux at 870~\mum\ and Bolocam flux at 1.1~mm are only upper limits. The continuous line shows the best-fit model obtained using the \cite{Robitaille2007} fitting tool. No other fit is adequate within the interval $\chi^2 - \chi^2_{best} < 3n$, where $n$ is the number of data points used in the fitting. In addition, no other model could be found by increasing this range by a factor of 1000.}
    \label{fig_sed_mho2147}
\end{figure}

Figure~\ref{fig_sed_mho2147} shows the complete SED of IRAS~17527, covering wavelengths between 3.6~\mum\ and 1.1~mm. We fitted this SED using the grid of models and the fitting tool of \citet{Robitaille2006,Robitaille2007}. This grid consists of 20 000 two-dimensional radiation transfer models of axisymmetric YSOs, with each model providing an emergent SED at ten viewing angles, and giving a total of 200 000 SEDs. The YSO models cover stellar masses from 0.1 to 50~\msun, and with evolutionary stages ranging from the very early stage of envelope infall to the late disc-only stage. The geometry of the YSO is determined by the combination of a central stellar source, disc, and  infalling envelope with bipolar cavities \citep{Whitney2003a,Whitney2003b}. The models are deterministic for parameters of the star, disc, and envelope, such as: stellar radius, masses, radii, and mass accretion rates of the disc and envelope.

The tool retrieves only one model fit, with the difference between its $\chi^2$ value and the best or minimum $\chi^2_{best}$ being less than $3n$, where $n$ is the number of data points used in the fitting \citep{Robitaille2007}. We increased this range ($\chi^2-\chi^2_{best}<3n$) by a factor of 1000, and even then, no other fit was within this larger interval. Table~\ref{tab_mho2147_sed} lists the corresponding parameters. \citet{Laws2019} found only one valid fit for the SEDs of two sources (\#1 and \#5) associated with IRDC~G79.3$+$0.3, suggesting that the corresponding models were well constrained by the available fluxes. \citet{Tapia2020} encountered a similar situation, in that a single model was found for the SED modelling of IRAS~12272$-$6240. Overall, our model for IRAS~17527, based on a better defined SED and an improved version of the SED fitting tool, agrees reasonably well with the one derived by \cite{Varricatt2011}, and is within the range of parameters obtained for other intermediate-mass and massive young stars of similar characteristics \citep[see e.g.][]{Fazal2008, Grave-Kumar2009,Lui2019}. 

In general, the Robitaille et al. model \citeyearpar{Robitaille2007} provides a good fit to the observed fluxes, by adjusting many free parameters at the expense of some physically unrealistic values or degeneracies \citep[see e.g.][]{DeBuizer2017,Jones2019}. This is probably the case too for the infalling envelope mass of IRAS~17527, which was found to be rather large. \citet{Guzman2010} suggested that the Robitaille et al.  model \citeyearpar{Robitaille2007} overestimates the mass of the envelopes by a factor of between two and three. Considering this overestimation, we note that the envelope mass derived from the SED modelling (2437~\msun) is roughly in agreement with that obtained by integrating the \H2 column density over the elliptical area in Fig.~\ref{fig_mho2147_dens} (1400~\msun, see Sect.~\ref{sec_mass-clump_mho2147}). In any case, it is worthwhile mentioning that, although this SED fitting tool provides a model that is consistent with the observed fluxes, it is unable to reproduce the detailed complexity of the true system. Thus, in some cases, the SED fitting tool does not provide a self-consistent set for a large number of free parameters.

\begin{table}
\centering
\begin{threeparttable}
    \caption{Physical parameters of IRAS~17527$-$2439 obtained from the SED fitting.}
    \label{tab_mho2147_sed}
    \begin{tabular}{ll}
    \hline
    \hline
    \noalign{\smallskip}
    Parameters & IRAS~17527$-$2439 \\
    \hline
    \noalign{\smallskip}
    Stellar age (yr)                & $3.3 \times 10^4$ \\
    Stellar mass (\msun)            & $11.2$ \\
    Stellar radius (\rsun)          & $27.2$ \\
    Stellar temperature (K)         & $9345$ \\
    Disc mass (\msun)               & $0.4 $ \\ 
    Disc accretion rate (\msun\ yr$^{-1}$)  & $1.9 \times 10^{-6}$ \\
    Disc inner radius (AU)          & $5.2$ \\ 
    Disc outer radius (AU)          & $30.5$ \\
    Envelope mass (\msun)           & $2437$ \\
    Envelope radius (AU)            & $1.0 \times 10^5$ \\
    Envelope accretion rate (\msun\ yr$^{-1}$)  & $3.6 \times 10^{-3}$ \\
    Total luminosity (\lsun)        & $5.1 \times 10^3$ \\
    A$_V$ (mag)                     & $38.3 $ \\
    Angle of inclination (\degr)    & $18$ \\
    \hline
    \end{tabular}
    \end{threeparttable}
\end{table}

\begin{figure*}
    \centering
    \includegraphics[width=\columnwidth, trim={0cm 0cm 1cm 0cm}, clip]{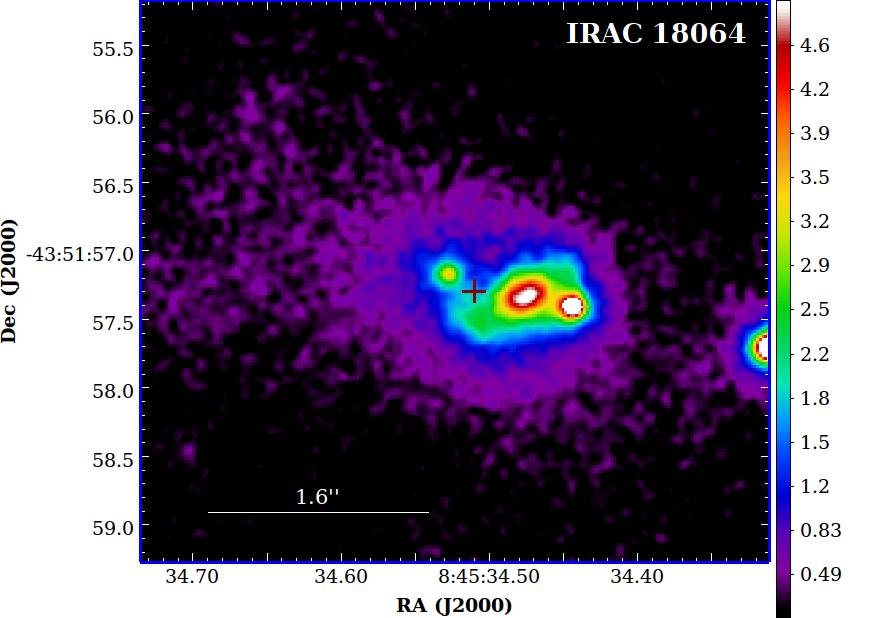}
    \includegraphics[width=\columnwidth, trim={0cm 0cm 2cm 0cm}, clip]{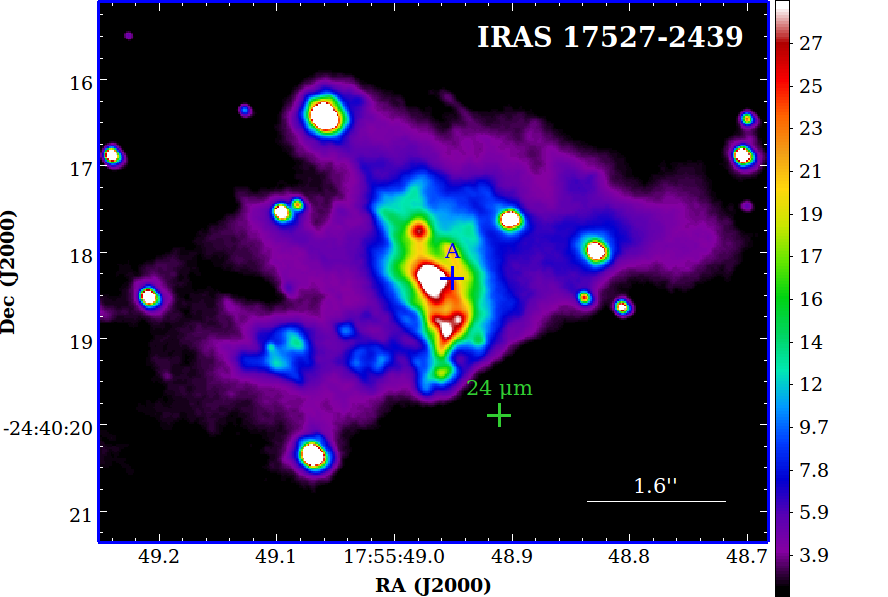}
    \caption{{\textit{Left panel:} High-resolution K-band image of the field around the proposed exciting sources of MHO~1502. The dark red cross indicates the position of IRAC~18064, which is associated with MHO~1502. \textit{Right panel:} High-resolution K-band image of the field around the suggested driving star of MHO~2147. The blue and green crosses mark the location of `source A' (only detected in the K and \H2 bands) and a bright source at 24~\mum, respectively. Both these sources were identified by \cite{Varricatt2011} and are related to IRAS~17527$-$2439. The white bar in both panels shows the FWHM of IRAC/\Spitzer\ \citep{Fazio2004} as a reference.  The flux scale is shown to the right and is calibrated in units of $10^{-8}$~Jy.}}
    \label{fig_sources_k_filter} 
\end{figure*}

\subsection{The possible multiplicity of the proposed exciting sources for MHO~1502 and MHO~2147}
\label{sec_multiplicity_exciting_sources}

The left and right panels of Figure~\ref{fig_sources_k_filter} show high-resolution K-band images of the fields where the proposed exciting sources for MHO~1502 (IRAC~18064) and MHO~2147 (IRAS~17527$-$2439), respectively, are located. In the case of MHO~1502, we detected two relatively bright objects close to the position of IRAC~18064 \citep{Giannini2013}, which are marked with a dark red cross in this figure. The differential magnitude between the two components $\Delta$K is $\sim 0.8$~mag, and the two stars are separated by 0.3\arcsec, $\sim 240$~AU at a distance of 700~pc. The elongated shape of the eastern component may even indicate that it is an unresolved double star. However, IRAC~18064 may not be the true exciting source of the MHO~1502 jet, because it is located away from the axis of the jet, as discussed below in Sect.~\ref{sec_models_masciadri}. 

With regard to MHO~2147, the blue and green crosses in the right panel of Fig.~\ref{fig_sources_k_filter} indicate the positions of source A (only detected in the K and \H2 bands) and a bright 24~\mum\ source, respectively, which are both associated with IRAS~17527$-$2439 \citep[][see Sect.~\ref{sec_exciting-source_mho2147}]{Varricatt2011}. Source A might be a triple system with a separation of 0.7\arcsec, or $\sim 2100$~AU (at a distance of 3.23~kpc) between the northern component and the two dim objects to the south. The difference in the K band is $\sim 0.7$~mag, obtained by measuring the two faint sources to the south together. These latter sources are separated by 0.18\arcsec, or $\sim 600$~AU, assuming the same distance. A fourth very faint object appears to the northeast of the source. 

In summary, our K-band image resolved the previously proposed exciting sources into two or more components, suggesting binary or multiplicity properties of the driving stars. However, a physical association between the stellar components needs to be confirmed. In this sense, high-angular resolution and sensitive infrared or submillimetre multiple-epoch images are required to measure proper motions or to determine the relative orbits.

\section{Jet models for MHO~1502, MHO~2147 and the Ad--jet}
\label{sec_models_masciadri}

\subsection{Description of the model}

To describe the meandering profiles of MHO~1502, MHO~2147, and the Ad--jet, we used a model developed by \cite{Masciadri2002}, in which the wiggling jet morphology can be produced by two scenarios, the orbital movement of a binary star or the precession of the jet axis.  Because stellar jets associated with young stars  are hypersonic and have higher densities than their immediate environment, they are modelled as being ballistic\footnote{In a ballistic jet, the fluid parcels preserve the velocity at which they are ejected.}.

The orbital model considers that the exciting source of the jet describes a circular orbit (contained in the $xy$-plane) of radius $r_0$, orbital period T$_0$, orbital velocity $v_{0}=2\pi r_0/{\rm T_{0}}$, and angular frequency $\omega=2\pi/{\rm T_0}$, assuming that the jet and the orbital axis are parallel. The shape of the jet projected onto the $xz$-plane can be expressed by:
\begin{equation} 
     \left(\frac{x}{r_0}\right) = ~\kappa \left(\frac{z}{r_0} \right)\sin{\left[{\kappa \left(\frac{z}{r_0}\right) - \omega t}\right]} + \cos{\left[{\kappa \left(\frac{z}{r_0}\right) - \omega t}\right]}, \label{eq_orbital} 
\end{equation}

\noindent
where $\kappa = \frac{v_{0}}{v_{j}}$ and $v_{j}$ is the ejection velocity. We assumed $v_j~=~100$~\kms\ \citep{Reipurth-Bally2001}. 

With regard to the precession effect, several causes have been proposed. One of these is a tidal effect exerted on the disc that surrounds the exciting source of the jet by the presence of a companion star whose orbit is not coplanar with the plane of the disc \citep{Raga1993,Terquem1999,Shepherd2000}.
Again, following the formalism of \cite{Masciadri2002}, the precession period is given by $T_p = 2\pi/\nu$, 
where $\nu$ is the frequency, and the trajectory of the jet can be represented as:
\begin{equation}
     x = z \tan{\beta} \cos{\left[\nu \left(t - \frac{z}{v_j \cos{\left(\beta\right)}}\right)\right]},
      \label{eq_precession}
\end{equation}

\noindent 
where $\beta$ is the half-opening angle of the precession cone.

We note that while the orbital model has a mirrored behavioural symmetry with respect to the orbital plane, the precession model has a point symmetry with regard to the location of the source. Therefore, the shape of a jet with a strong `S-type' wiggling (close to the exciting source) should be easier to model using precession. Another characteristic that distinguishes one model from the other is that, while the precession model must pass through $z=0, x=0$, this is not necessarily true for the orbital model. In this case, at $z=0$, $\|x\|$ can take values between $0\leq\|x\|\leq r_{0}$.

\subsection{The models applied to MHO~1502, MHO~2147, and Ad--jet}

The $\chi^2$ test was used to identify the model and the set of parameters that best reproduced the observed positions of the knots, and  we applied the Levenberg--Marquardt algorithm to minimise
the $\chi^2$ values. The positions of the knots along the MHO~1502 and those identified as belonging to the Ad--jet were measured on the \H2 image, with the actual position corresponding to the brightest pixel of each knot. In the case of MHO~2147, and to better delineate the jet extension and terminal region, many consecutive points along the axis jet were measured. To determine the centre of the modelled jet, in each case we tested the positions of all individual sources for the different wavelengths associated with the proposed exciting sources (see Sects.~\ref{sec_MHO1502} and \ref{sec_exciting-source_mho2147}). In the case of MHO~1502, none of the known sources are perfectly aligned with the jet axis. Consequently, we explored several positions, such as those of IRAC~18064, as well as several locations along the axis of the jet, including the position of a core of dust continuum at 1.2~mm and identified as MMS2 by \cite{Massi2007}. For MHO~2147 and the Ad--jet, the position of the point source at 24~\mum\ detected by \citet[][see Sect.~\ref{sec_exciting-source_mho2147}]{Varricatt2011} gives the best fits\footnote{We used the position of the source at 24~\mum\ as the centre of both jets, although as discussed in Sect.~\ref{sec_MHO2147}, the driving source of each jet is likely to be different.}.

Table~\ref{tab_models} lists the parameters obtained for both models (binary orbital motion and precession) for each jet. Given the extension and appearance of the jets in the images, and assuming that the jets roughly coincide with the plane of the sky, we adopted an inclination angle of $i = 90\degr$ (see Figs.~\ref{fig_MHO1502} and \ref{fig_MHO2147}). Inclination angles that are not 90\degr\ affected the parameters listed in this table, in particular the orbital period in the case of the orbital model and the opening angle for the precession scenario. In addition, the orbital radius in the orbital model was not significantly affected by the inclination angle. In any case, for $i > 70\degr$, the variations of the parameters listed in Table~\ref{tab_models} are within the expected errors. Inclination angles $<70\degr$ are very unlikely considering the extensions of jets.

Figure~\ref{fig_curvas1} shows the orbital and precession models for MHO~1502 obtained by choosing a centre along the jet axis at \radec~$=$~(08:45:34.9; $-$43:51:55.9) (red and magenta lines), and at the position of IRAC~18064 (dotted grey lines) superimposed on the combined \H2 (green) and K (magenta) band images. The position of the centre along the jet axis is displaced $\sim2064$~AU (or $\sim4$\arcsec) to the northeast from IRAC~18064 (see Fig.~\ref{fig_curvas1}). Both the orbital and precession models centred on the jet axis reproduced the jet profiles
satisfactorily, although the orbital model yielded a slightly smaller $\chi^2$ value (see Table~\ref{tab_models}).

Regarding the models centred on IRAC~18064, the orbital model has a smaller $\chi^2$ than the precession model. The most likely reason for this is that the former model is not constrained to passing through $z=0$, $x=0$, and can therefore shift towards the jet axis, resulting in a better fit. Moreover, we note that the displacement between the jet axis and IRAC~18064 ($\sim 2064$~AU) is roughly compatible with the orbital radius ($\sim 2153$~AU) for the orbital model centred on this source (see Table~\ref{tab_models}). This is also a consequence of the model: near $z=0$, $x=0$ the knots in the jet cannot travel much farther than the orbital radius (depending on the ejection and orbital velocities). Therefore, this model (with the centre on IRAC~18064) would require the existence of a companion star to IRAC~18064 at a separation similar to the distance between the knots close to $z=0$ and IRAC~1806 and at a distance of $\sim 2000$~AU.

Additionally, the models centred on IRAC~18064 have larger residuals and larger $\chi^2$ values than those centred on the jet axis. This would favour the existence of an as-of-yet unidentified exciting source located along the jet axis. The requirement of the precession model to pass through the origin of the coordinate system (i.e. at the location of the exciting source) also supports this scenario, suggesting that IRAC~18064 would not be the exciting source for the jet MHO~1502 (if we accept precession as an explanation for the observed shape of the jet). However, in the case of the orbital model, the scenario is more complex given that this model is not constrained to passing through the centre of the coordinates. According to the resulting $\chi^2$ values in Table~\ref{tab_models}, the orbital model also seems to favour the centre on the jet axis, but the difference between the $\chi^2$ values for both centers is not sufficient enough to completely discard IRAC~18064 as the possible centre. As mentioned above, a companion at a separation of $\sim~2153$~AU from IRAC~18064 may explain the observed shape of the jet, although this companion is not compatible with the nearby star (at $\sim~240$~AU) seen in the left panel of Fig~\ref{fig_sources_k_filter}.

In any case, the best models in Table~\ref{tab_models} (orbital and precession) are the ones that are centred on the jet axis. The relatively large separation of the companions in the orbital model centred on IRAC~18064 casts some doubt on the correct identification of IRAC~18064 as the MHO~1502 driving star, while it suggests the centre of the jet as being the best place to search for an alternative exciting source. The orbital model centred on the jet axis predicts a binary star with a separation of $\sim 550$~AU (see Table~\ref{tab_models}). 

In the case of MHO~2147 and the Ad--jet, the binary orbital motion does not provide acceptable fits, as can be inferred from the $\chi^2$ values listed in Table~\ref{tab_models}. Figure~\ref{fig_curvas2} displays the best fits for the precession models for MHO~2147 (continuous red line) and the Ad--jet (cyan squares). \cite{Varricatt2011} suggests a precession scenario for MHO~2147 based on a slight counter-clockwise rotation of the 4.5~\mum\ and K-band emissions in relation to the \H2\ band, particularly in the SW lobe.

\subsection{The parameters of the models}

\begin{figure*}
        \centering
    \includegraphics[width=0.5\columnwidth]{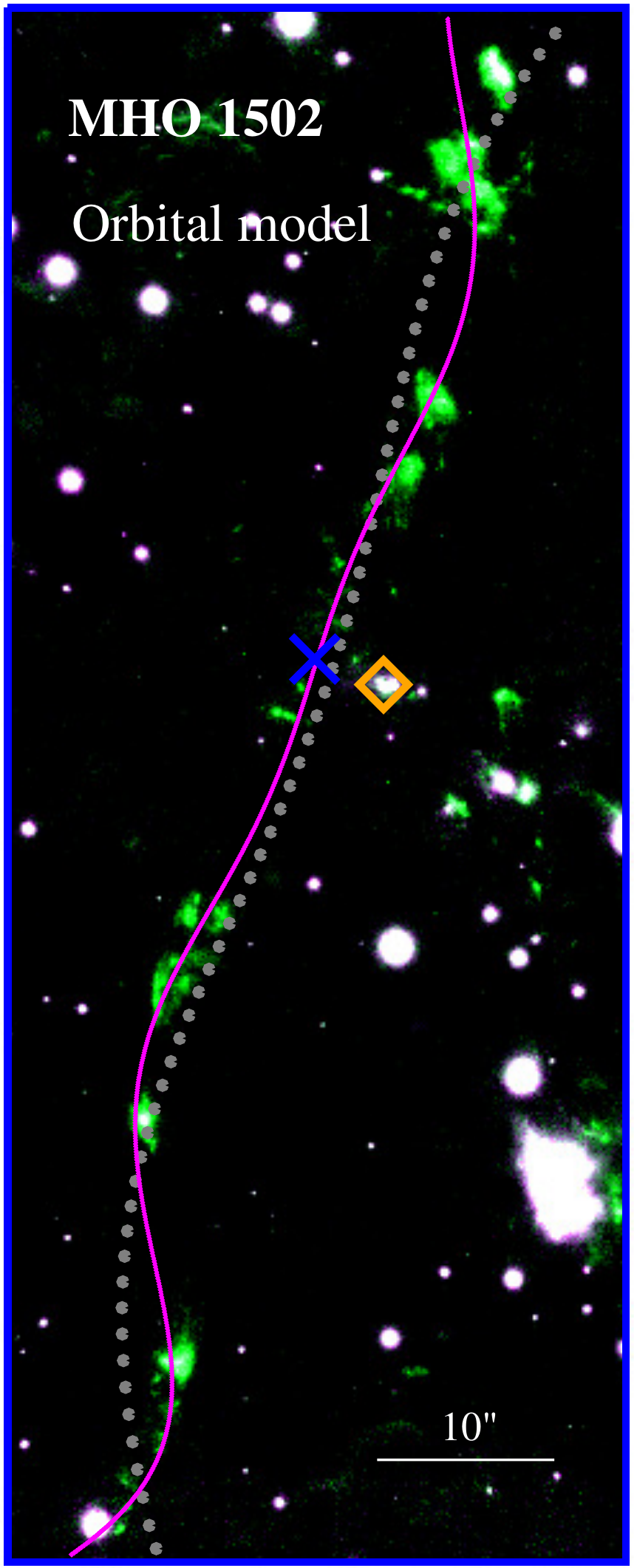}
    \includegraphics[width=0.5\columnwidth]{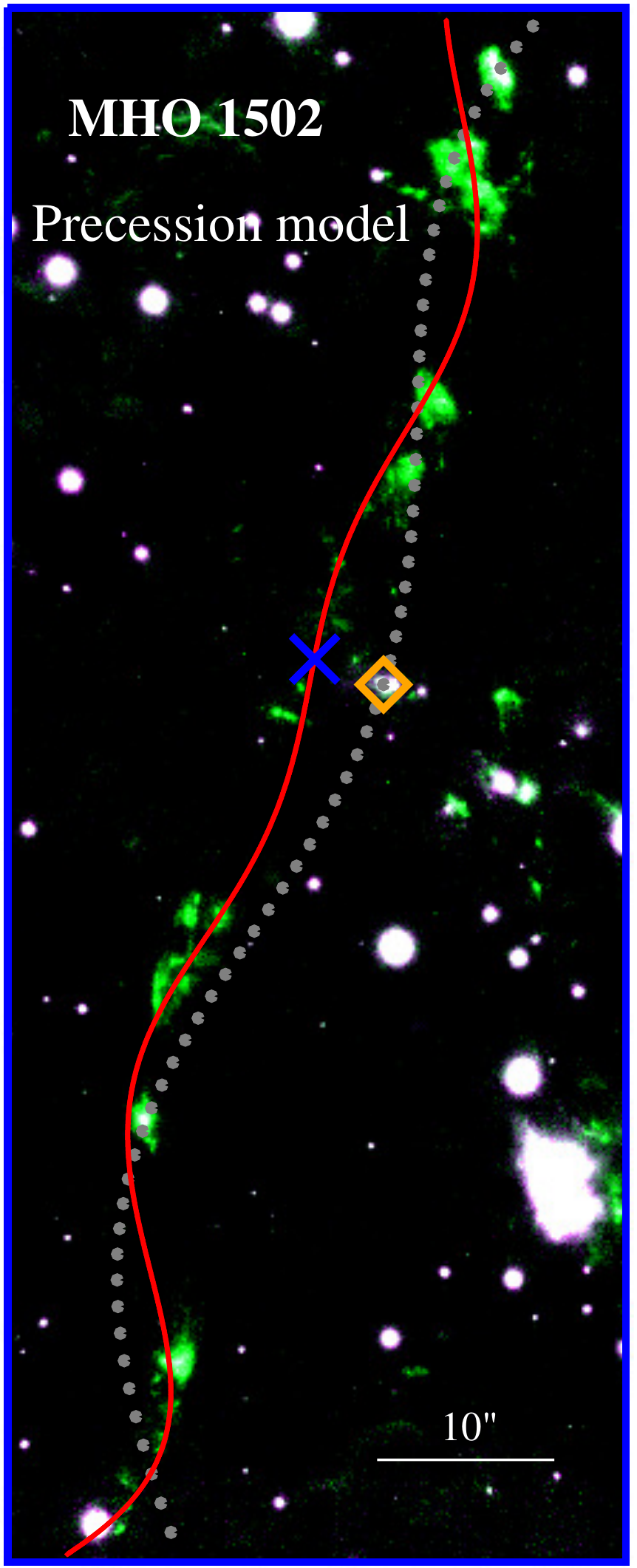}
        \caption{Orbital (left panel) and precession (right panel) models for MHO~1502 superposed on the combined \H2 (green) and K (magenta) band image. The blue cross at \radec~$=$~(08:45:34.9; $-$43:51:55.9) indicates the centre along the jet axis that provides the best fits, and the orange diamond marks the position of IRAC~18064 identified by \citet{Strafella2010}. The magenta and red lines show the models centred on the blue cross. The dotted grey lines are the models centred on the IRAC~18064 source.}
        \label{fig_curvas1}
\end{figure*}

\begin{figure*}
        \centering
    \includegraphics[width=0.7\textwidth]{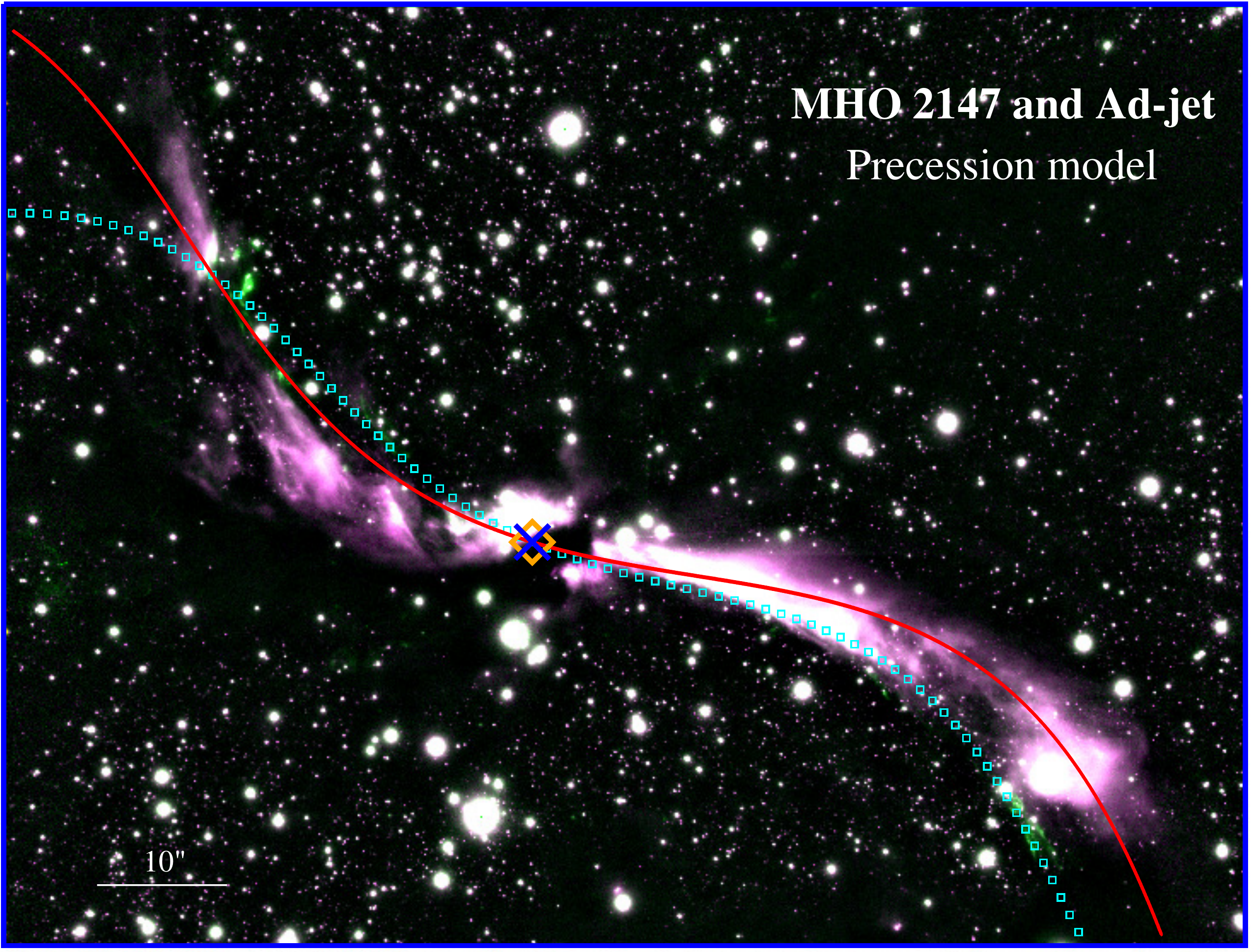}
        \caption{Precession models for MHO~2147 (continuous red line) and for the Ad--jet (cyan squares) superposed on the combined \H2 (green) and K (magenta) band image. The centre of the jet models (blue cross) coincides with the position of a 24~\mum\ source (orange diamond) detected by \citet{Varricatt2011}.}
    \label{fig_curvas2}
\end{figure*}

Assuming that Fig.~\ref{fig_curvas1} shows the total length of the MHO~1502 jet, by adopting a velocity of 100~\kms, we derived a dynamical time of $2.9\times10^3$~yr. If we compare this time with the orbital ($T_o$) and precession (T$_p$) periods in Table~\ref{tab_models}, we find that the driving source has completed two revolutions in the case of the model centred on the jet axis, but only one period for the model centred on IRAC~18064. We also note that the orbital and precession periods are consistent with the age of the stellar cluster IRS~16 and the dynamical age of the \HII\ region G263.619$-$0.533 \citep{Massi2003,Massi2010}. Similarly, for MHO~2147 and the Ad--jet (considering that Fig.~\ref{fig_curvas2} shows the complete extensions of the jets), the dynamical times are $\sim 7.4\times10^3$~yr and $\sim 5.5\times10^3$~yr, respectively. By comparing these values with the corresponding T$_p$ (see Table~\ref{tab_models}), we can infer that the driving source of MHO~2147 has completed approximately three precession cycles, while the exciting source of the Ad--jet has realised approximately two precession cycles. The dynamical age of MHO~2147 is roughly consistent with the stellar age ($\sim$ $ 3.3\times10^4$~yr) of the proposed exciting star, IRAS~17527$-$2439, derived from the SED modelling (see Table~\ref{tab_mho2147_sed}). In addition, the precession periods derived for MHO~2147 and the Ad--jet are within the estimated range for MYSOs \citep{Frank2014}.

In general, the parameters listed in Table~\ref{tab_models} are similar to others determined for different jets associated with YSOs in different star-forming regions. For example, \cite{Lee2010}, \cite{Noriega-Crespo2011}, and \cite{Estalella2012} applied the binary orbital motion to the well-known jets HH~211, HH~111, and HH~30. The precession scenario has been applied to the L1157 jet \citep{Gueth1996, Podio2016}, the outflow driven by Cep~E~A and associated with HH~377 \citep{Eisloffel1996}, the HH~34 jet \citep{Devine1997}, the jet driven by the IR source UGPS~J185808.46$+$0100041.8 \citep{Paron2016}, and the V380 Ori NE jet \citep{Choi2017}. In other cases, both models (i.e. binary orbital motion and precession) reproduced the observed shape well, with no significant differences between them. \cite{Anglada2007} arrived at such a conclusion when analysing the HH~30 jet and counter-jet. Conversely, for the HH~47 jet, \citet{Reipurth2000} argued that neither the binary orbital motion nor the precession models are compatible with the observed characteristics of this jet.

In summary, the best orbital motion model for MHO~1502 indicates an as-of-yet undetected driving source located on the jet axis in a relatively obscure region with no known sources (see the insert in the left panel of Fig.~\ref{fig_mho1502_hii_spitzer}). On the other hand, the binary model centred on IRAC~18064 (the proposed driving star located off the jet axis) suggests the existence of a binary companion at a distance not compatible with the nearby star seen in the left panel of Fig~\ref{fig_sources_k_filter}. High spatial- and spectral resolution molecular line observations, such as those obtained by the ALMA telescope, can provide information on the surrounding environment of the unknown driving source(s) \citep{Motte2018}, although IR/submm data would provide a means of possibly obtaining a better characterisation of such exciting sources. For MHO~2147 and the Ad--jet, the precession model provides good fits.

\subsection{MHO~2147, Ad--jet, and MHO~2148: a triple jet system?}

MHO~2147 may be associated with the Ad--jet as well as the MHO~2148 perpendicular jet (see Fig.~\ref{fig_MHO2147}). Our precession models for MHO~2147 and the Ad--jet share a common centre and they fit  the S-shape profiles of both jets reasonably well. The much fainter MHO~2148, an almost perpendicular jet, also seems to originate from the same centre as MHO~2147 and the Ad--jet, hinting at a multiplicity of exciting sources. In this sense, the three jets might form a multiple jet system driven by nearby or gravitationally bound stars. Our high-resolution K-band image (see Fig.~\ref{fig_sources_k_filter} and Sect.~\ref{sec_multiplicity_exciting_sources}) shows a multiple stellar system that coincides with the position of source A and is associated with IRAS~17527$-$2439, the proposed driving star of MHO~2147 \citep{Varricatt2011}.

The MHO~2147/Ad--jet pair might be similar to a pair of jets in the massive star-forming region of Cepheus A: namely a pulsed precessing jet driven by a massive star (HW2) in a binary system, and a second outflow driven by a nearby source (HW3c or HW3d). The former jet, driven by HW2, has changed its orientation by about 45\degr\ in $\sim 10^4$ years and is associated with HH~174 and HH~169. This latter jet is located adjacent to the main jet (powered by HW2) on the plane of the sky, with HH~168 powering the westernmost lobe of this secondary outflow \citep{Cunningham2009,Zapata2013}. G35.20$-$0.74N is another example of multiple cores and nearby jets in a massive star-forming region and harbours three  cores, A, B, and C, with core B being composed of two sources. In addition, binary core B is the driving source of a precessing jet in the NE–SW direction and a radio jet, while core A is probably associated with a second jet lying in an almost E–W direction \citep{Gibb2003, Caratti_Garatti2015,Beltran2016}.

MHO~2147 and the almost perpendicular jet MHO~2148 (see Figs.~\ref{fig_MHO2148_AB}, \citealt{Varricatt2011,Ferrero2015a}) might be similar to other nearly orthogonal jets, such as the well-known HH~111/HH~121\citep{GredelReipurth1993,Reipurth2000,Rodriguez2008,Sewilo2017} and the outflows associated with Cep~E~A (HH 377) and Cep~E~B \citep{Terquem1999, Ospina-Zamudio2018}.

The similarity of MHO~2147, the Ad--jet, and the perpendicular MHO~2148 jet to other  previously reported jets (such as Cep~A, G35.20$-$0.74N, HH~111/HH~121, and Cep E A/Cep E B) suggests the existence of a small but interesting group of adjacent and perpendicular jets that are interrelated and are likely to be associated. However, to shed light on the physical relation of MHO~2147, Ad--jet, and MHO~2148,  high-angular-resolution and sensitive multi-wavelength data are needed.


  \begin{table*}
    \centering
    \begin{threeparttable}
    \caption{Parameters obtained from the orbital and precession models for the jets MHO~1502, MHO~2147, and the Ad--jet.}
    \label{tab_models}
    \begin{tabular}{llD{+}{\,\pm\,}{-1}D{+}{\,\pm\,}{-1}D{+}{\,\pm\,}{-1}D{+}{\,\pm\,}{-1}} 
      \hline
      \hline
      \noalign{\smallskip}
    \multirow{3}{4em}{Model}    & \multirow{3}{4em}{Parameters}         & \multicolumn{2}{c}{MHO~1502} & \multicolumn{2}{c}{MHO~2147} \\
             &               & \multicolumn{1}{c}{Centred on} & \multicolumn{1}{c}{Centred on} & \multicolumn{1}{c}{Main jet} & \multicolumn{1}{c}{Adjacent jet} \\
            &               & \multicolumn{1}{c}{the jet axis}      &  \multicolumn{1}{c}{IRAC 18064}         & \multicolumn{1}{c}{\,}  & \multicolumn{1}{c}{\,}   \\
      \hline
      \noalign{\smallskip}
\multirow{4}{4em}{Orbital}      & \text{$r_0$ (UA)} & 551+46    & 2153+192  & 5875+1129  & 7895+1651       \\
                        & \text{$T_0$ (yr)}                     & 1464+30   & 3349+146  & 8167+460  & 14772+1529  \\
                        & \text{$v_0$ (km s$^{-1}$)}    & 11.2+0.2  & 19.2+0.4  & 21.5+0.9     & 15.9+0.8 \\
                        & \text{$\chi^2$}              & \multicolumn{1}{c}{6.6}                & \multicolumn{1}{c}{18.3}        & \multicolumn{1}{c}{149}         & \multicolumn{1}{c}{55}      \\
      \hline 
      \noalign{\smallskip}
\multirow{3}{4em}{Precession} & \text{$\beta$ (\degr)} & 6.6+0.7        & 12+2  & 17+1 & 12+1 \\
                        & \text{$T_p$ (yr s$^{-1}$)}            & 1428+30       & 3065+142  & 21666+747 & 14960+712 \\
                        & \text{$\chi^2$}                           & \multicolumn{1}{c}{9.3}   & \multicolumn{1}{c}{34.4}  & \multicolumn{1}{c}{27}    & \multicolumn{1}{c}{14}  \\
      \hline
    \end{tabular}
    \end{threeparttable}
  \end{table*}

\section{Summary and conclusions}
\label{sec_conclusions}

In this work, we present high-resolution \H2 images of two wiggling stellar jets MHO~1502 and MHO~2147, obtained with GSAOI$+$GeMS. The MHO~1502 jet is composed of a chain of knots delineating a wiggling profile, while MHO~2147 displays a gentle continuous emission in \H2 and shows an S-shape symmetry. In addition, our image of the field around MHO~2147 revealed two other jets: the faint MHO~2148, which lies almost perpendicular to MHO~2147 in the plane of the sky and previously reported by \cite{Varricatt2011}; and a chain of knots in \H2 slightly offset from the axis of MHO~2147 but otherwise following the trace along this jet axis and identified as the adjacent jet and referred to here as the Ad--jet. These three jets seem to share a common origin, hinting at the multiplicity of the driving source.

We used a model developed by \cite{Masciadri2002} to reproduce the wiggling profiles of MHO~1502, MHO~2147, and the Ad--jet. For MHO~1502, the models that provide the best fits are centred on the jet axis but displaced $\sim 2063$~AU (or $\sim4$\arcsec) to the northeast of IRAC~18064, the proposed exciting source, thereby casting doubt on the correct identification of IRAC~18064 as the driving source. Related to this, the orbital model centred on the axis predicts a binary star with a separation of $\sim 550$~AU as an alternative exiting source. On the other hand, the orbital model centred on IRAC~18064 suggests the existence of a binary companion at a distance of $\sim 2000$~AU from IRAC~18064, though not yet detected and incompatible with the nearby source seen in the left panel of Fig.~\ref{fig_sources_k_filter}. Nevertheless, this last model cannot be discarded based on the quality of the fit (see Fig.~\ref{fig_curvas1} and Table~\ref{tab_models}). 

Orbital motion and/or precession of the jet exciting source reproduce the wiggling profile morphologies of MHO~1502 and MHO~2147 and indicate a probable binarity of the driving sources. Gemini high-resolution K-band images reveal close companions to each of the suggested exciting stars, although, as mentioned above, our best model for MHO~1502 suggests an as-of-yet undetected driving (binary) source on the axis of the jet.

We also investigated the origin of many \H2 emissions that lie in the field of MHO~1502, but are not directly related to the jet (see Fig.~\ref{fig_MHO1502}). The emissions in the upper left corner of the figure are probably created by UV fluorescence photons originating from the nearby bipolar \HII\ region G263.619$-$0.533, while those in the middle and lower parts of the field are likely to be shock excited regions from YSOs in the vicinity of the IRS~16 cluster. 

In the case of MHO~2147, we derived the mass of the IRDC in which the exciting source IRAS 17527$-$2439 is embedded and modelled the SED of this source. A mass of $\sim 1400$~\msun\ for the IRDC and typical parameters of intermediate- and high-mass Class~I objects were obtained for the IRAS source.

\begin{acknowledgements}
We thank Dr. Rodrigo Carrasco, the GSAOI assisting manager, for suggestions during the observations and Dr. Mischa Schirmer for his support and advice with the THELI reduction process. We are also grateful to Dr. Alberto Petriella for his comments and suggestions.
This work is based on observations obtained at the Gemini Observatory and the Program ID: GS-2014A-Q-29, which is operated by the Association of Universities for Research in Astronomy, Inc., under a cooperative agreement with the NSF on behalf of the Gemini partnership: the National Science Foundation (United States), the National Research Council (Canada), CONICYT (Chile), Ministerio de Ciencia, Tecnolog\'{i}a e Innovaci\'{o}n Productiva (Argentina), Minist\'{e}rio da Ci\^{e}ncia, Tecnologia e Inova\c{c}\~{a}o (Brazil), and Korea Astronomy and Space Science Institute (Republic of Korea).
\end{acknowledgements}

\bibliographystyle{aa} 
\bibliography{biblio}

\begin{appendix}

\section{Enlarged images of MHO~1502}
\label{appendix_mho1502}
\begin{figure*}
 \centering
 \includegraphics[width=\columnwidth]{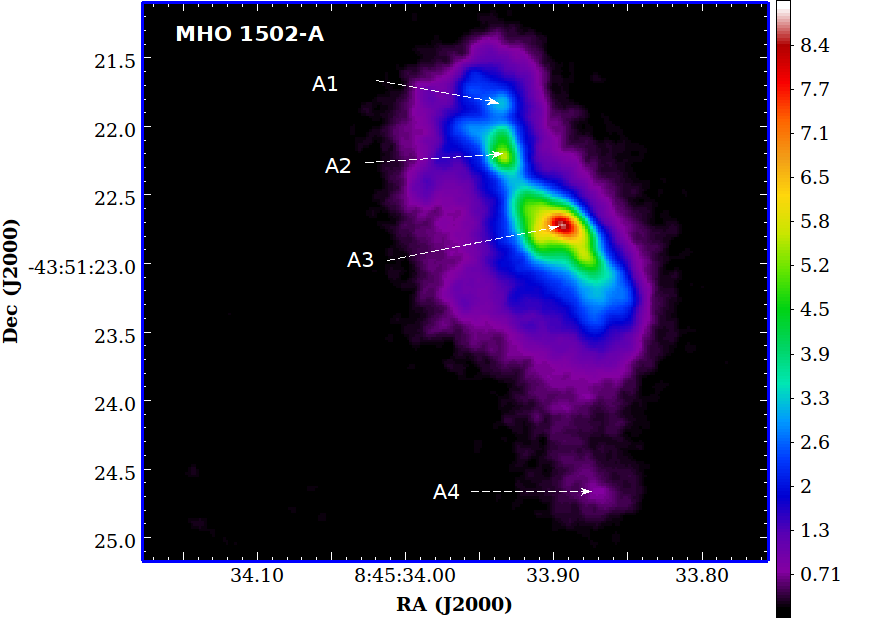}
 \includegraphics[width=\columnwidth]{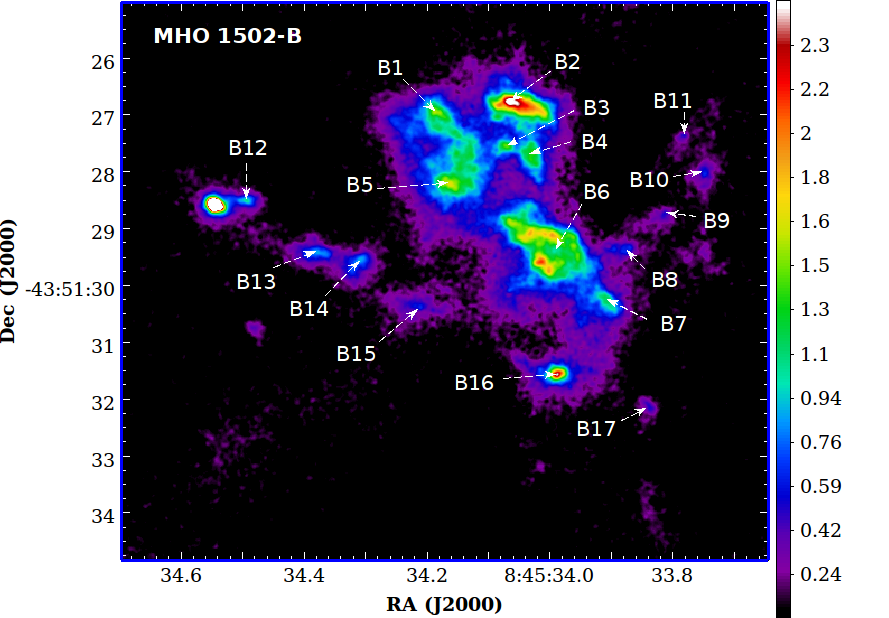}
 \par
 \hspace{0.1cm}
 \par 
 \includegraphics[width=\columnwidth]{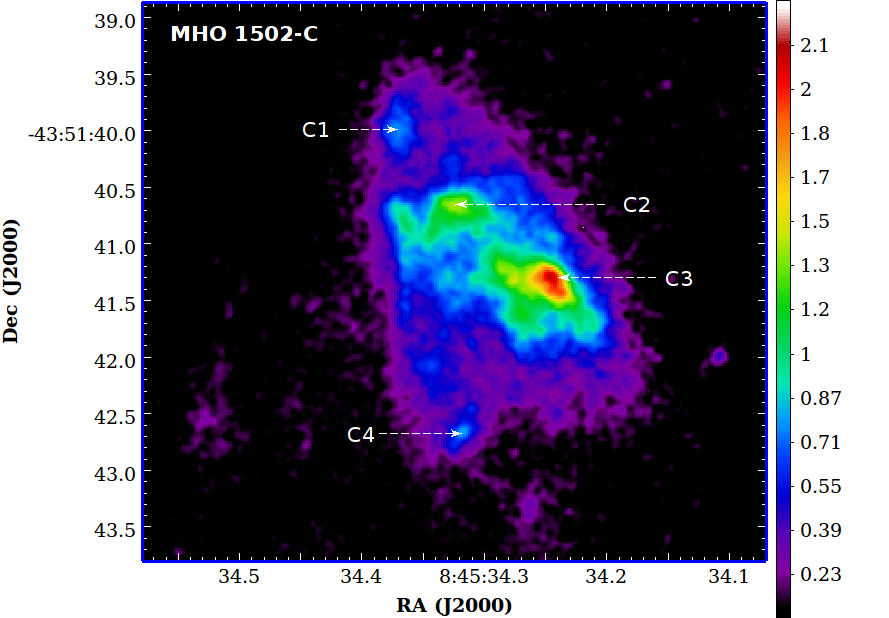}
 \includegraphics[width=\columnwidth]{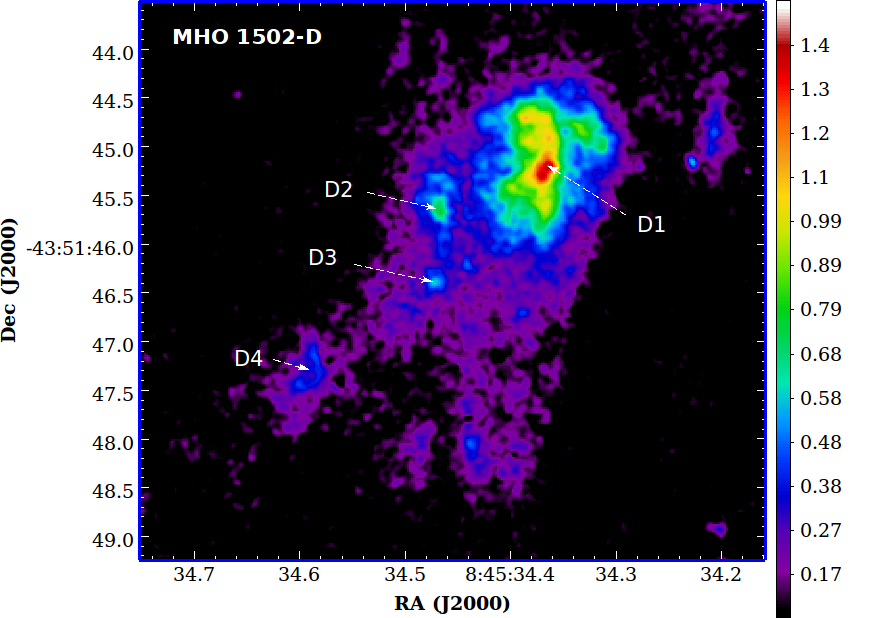}
 \par
 \hspace{0.1cm}
 \par 
 \includegraphics[width=\columnwidth]{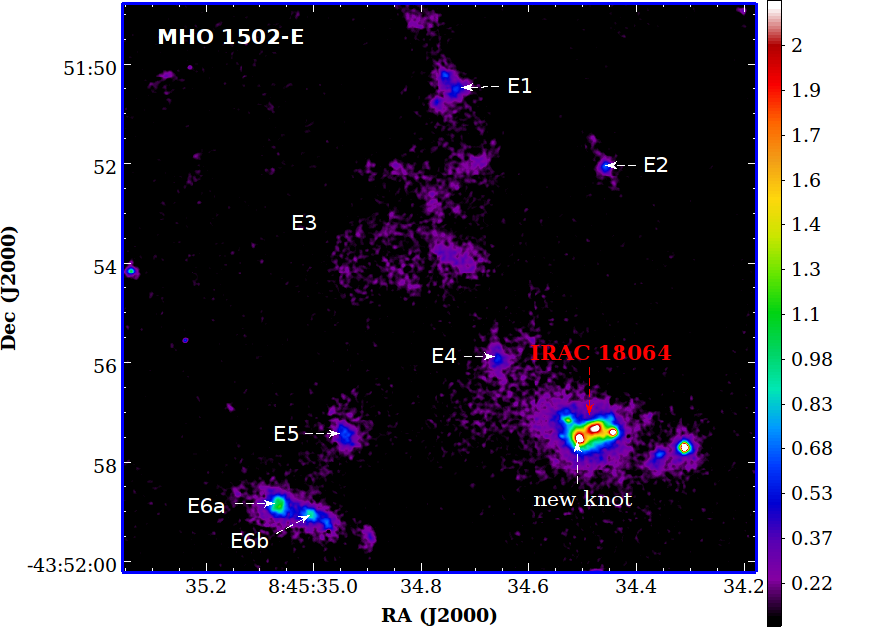}
 \includegraphics[width=\columnwidth]{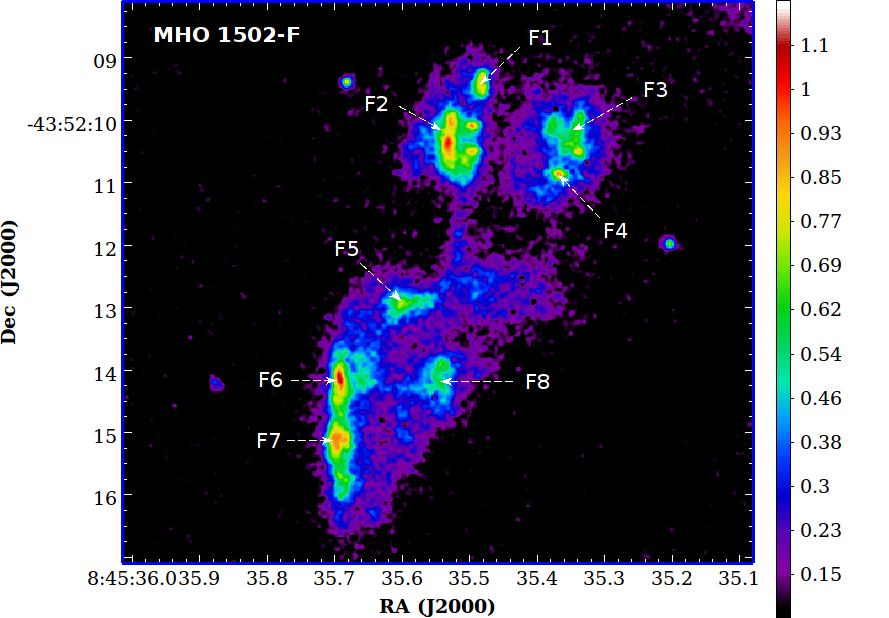}
 \caption{Knots MHO~1502-A (upper left panel), B (upper right panel), C (middle left panel), D (middle right panel), E (lower left panel), and F (lower right panel) in \H2 emission. The suggested exiting source by \cite{Giannini2013}, IRAC~18064, is indicated (in red) in the lower left panel. The flux scale is shown on the right and is calibrated in units of 10$^{-8}$~Jy.}
 \label{fig_MHO1502_ABCDEF}
\end{figure*}


\begin{figure*} 
 \centering
 \includegraphics[width=\columnwidth]{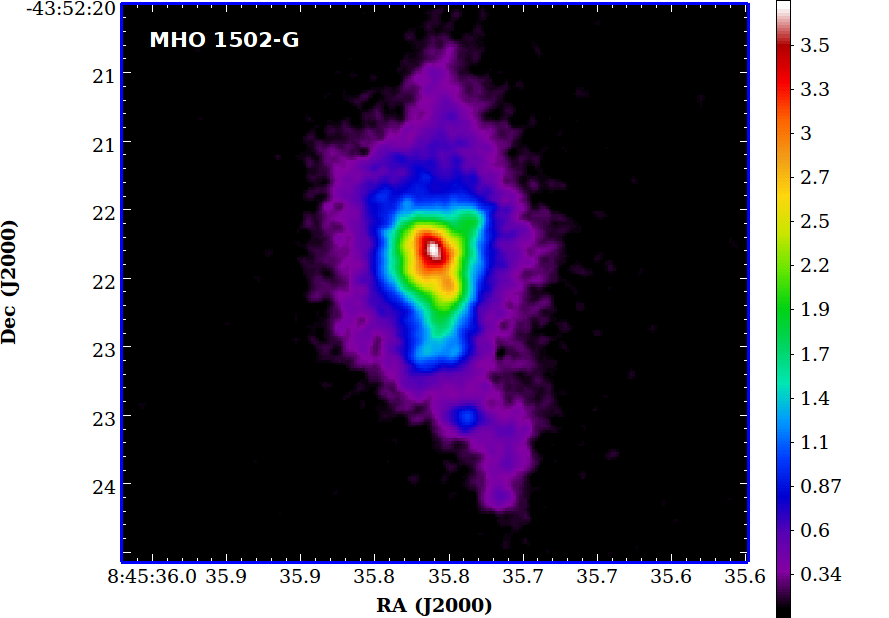}
 \includegraphics[width=\columnwidth]{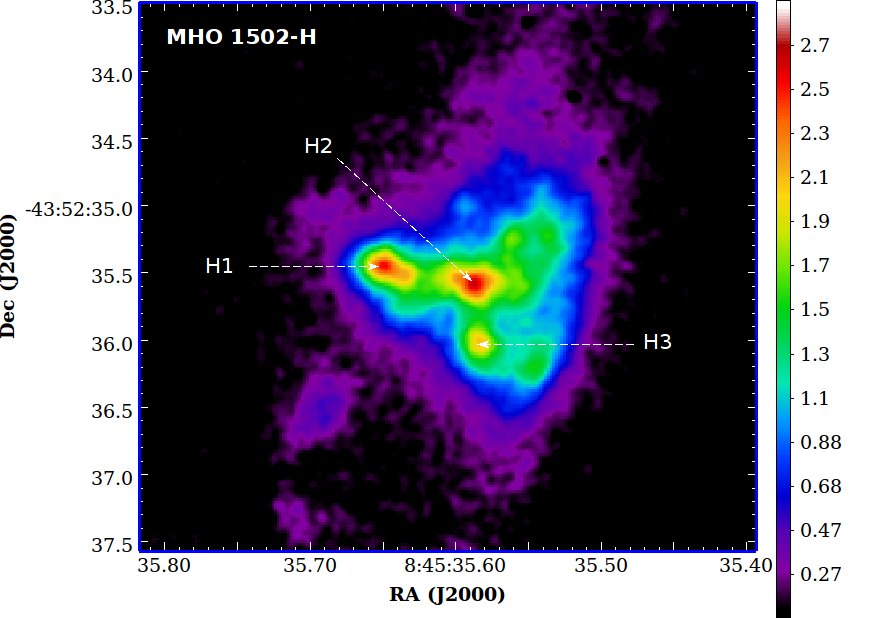} 
 \par
 \hspace{0.1cm}
 \par
 \includegraphics[width=\columnwidth]{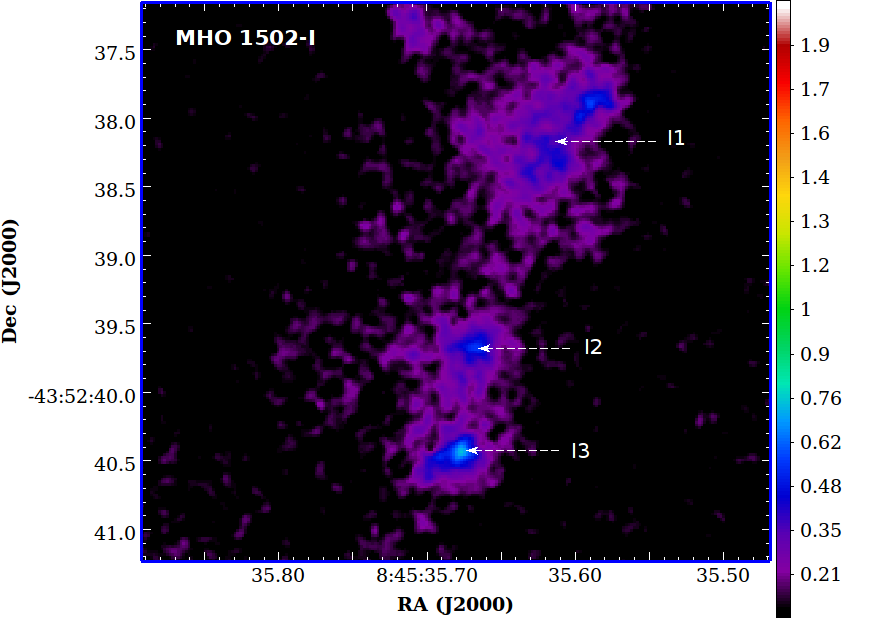}
 \includegraphics[width=\columnwidth]{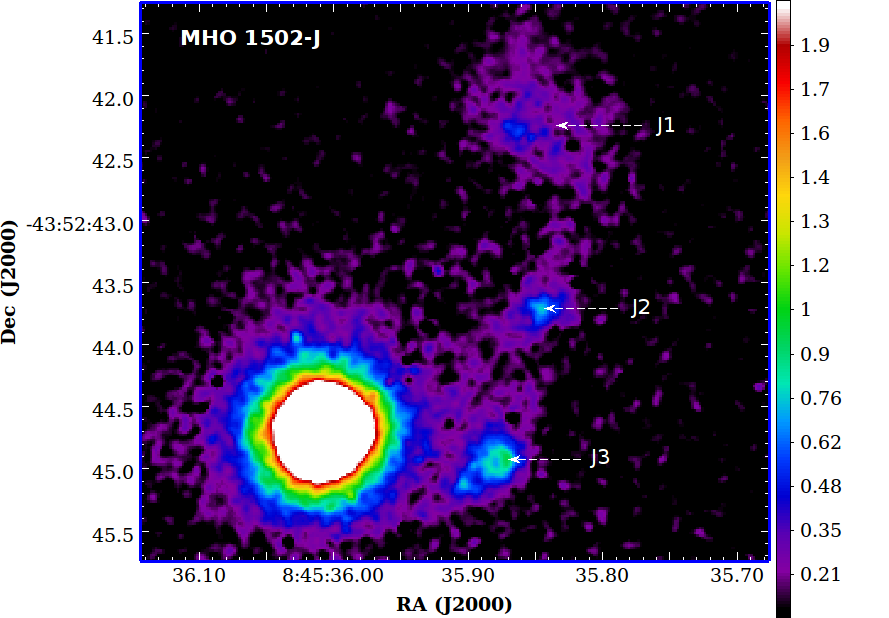}  
 \caption{Knots G (upper left panel), H (upper right panel), I (lower left panel), and J (lower right panel) in \H2 emission. The flux scale is shown on the right and is calibrated in units of 10$^{-8}$~Jy.}
 \label{fig_MHO1502_GHIJ}
\end{figure*}


\clearpage 
\newpage
\section{Enlarged images of MHO~2147 and the Ad--jet}
\label{appendix_mho2147}

\begin{figure*} 
  \centering
 \includegraphics[width=\columnwidth]{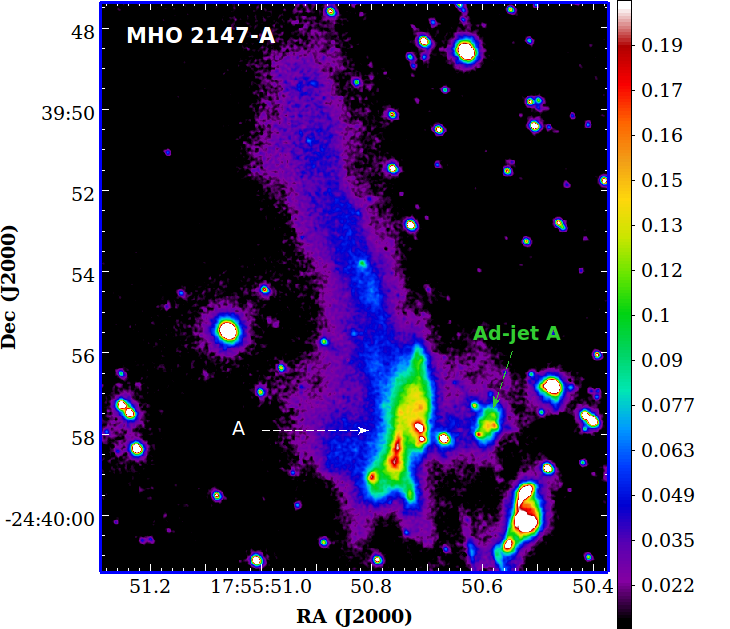}
 \includegraphics[width=\columnwidth]{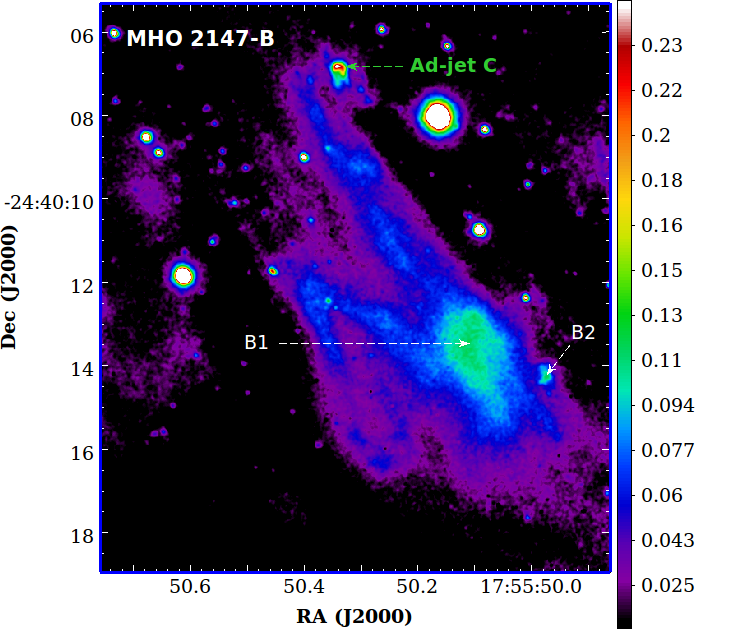}
 \par
 \hspace{0cm}
 \par
 \includegraphics[width=\columnwidth]{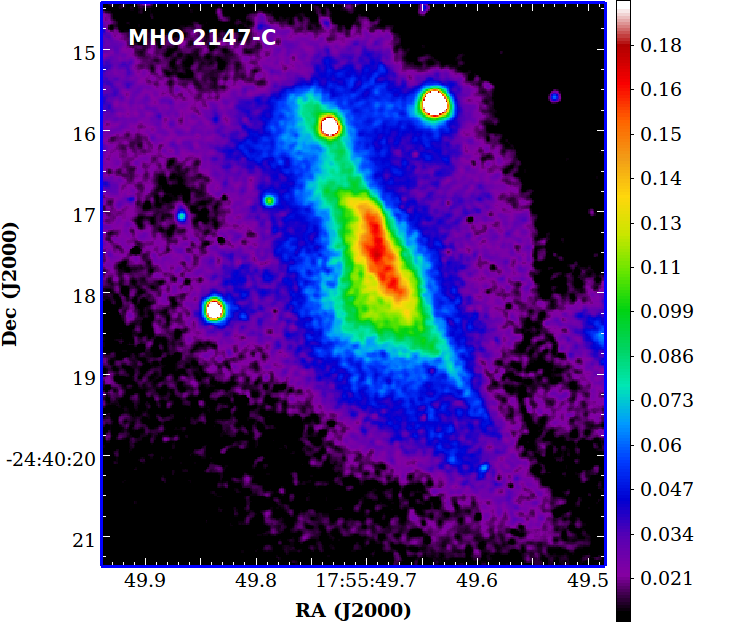}
 \includegraphics[width=\columnwidth]{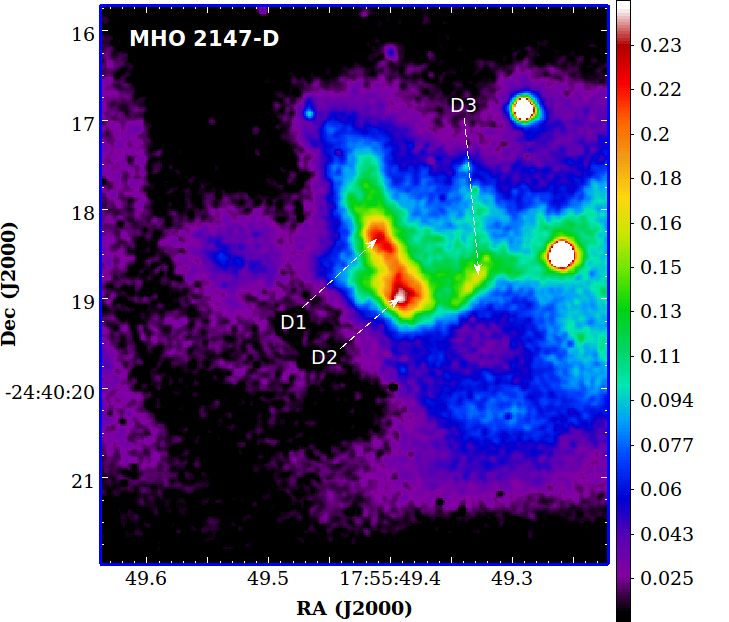}
 \par
 \hspace{0cm}
 \par
  \includegraphics[width=\columnwidth]{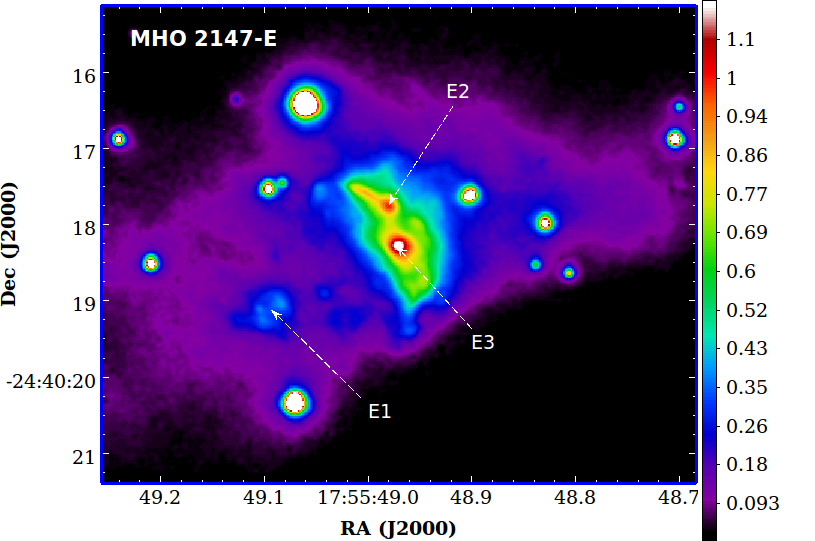}
  \includegraphics[width=\columnwidth]{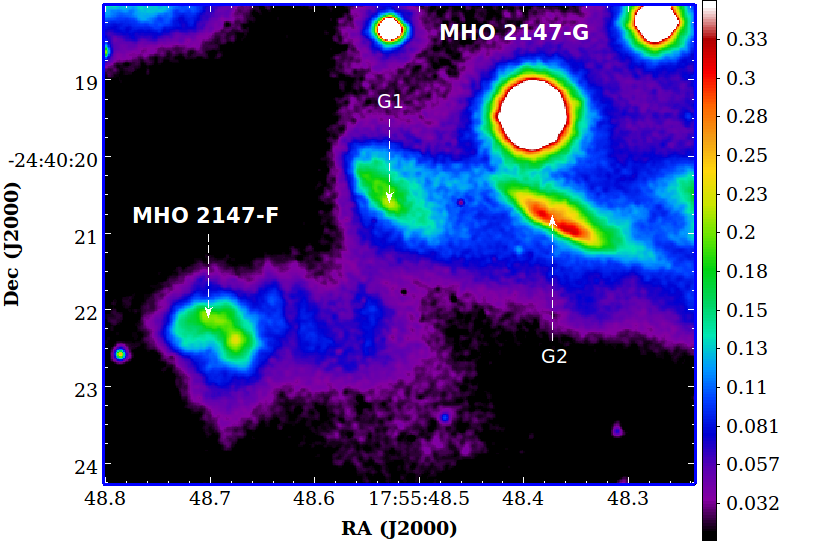}
 \caption{Knots MHO~2147-A (upper left panel), B (upper right panel), C (middle left panel), D (middle right panel), E (lower left panel), F, and G (lower right panel) in \H2 emission are indicated with dashed-white arrows. The dashed-green lines in the upper left and right panels indicate the knots A and C, respectively, which seem to belong to the Ad--jet. The flux scale is shown on the right and is calibrated in units of 10$^{-8}$~Jy.}
 \label{fig_MHO2147_ABCDEFG}
\end{figure*}


\begin{figure*}
    \centering
    \includegraphics[width=\columnwidth]{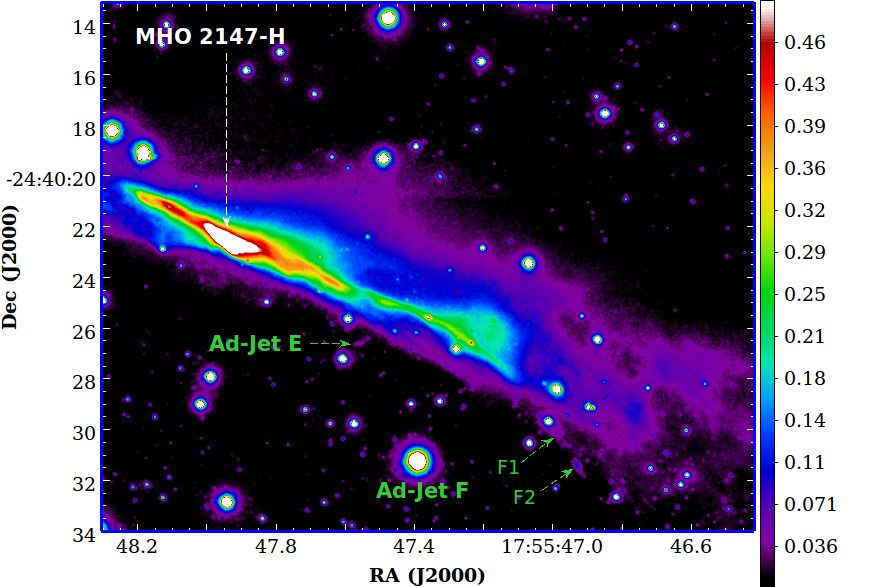}
  \includegraphics[width=\columnwidth]{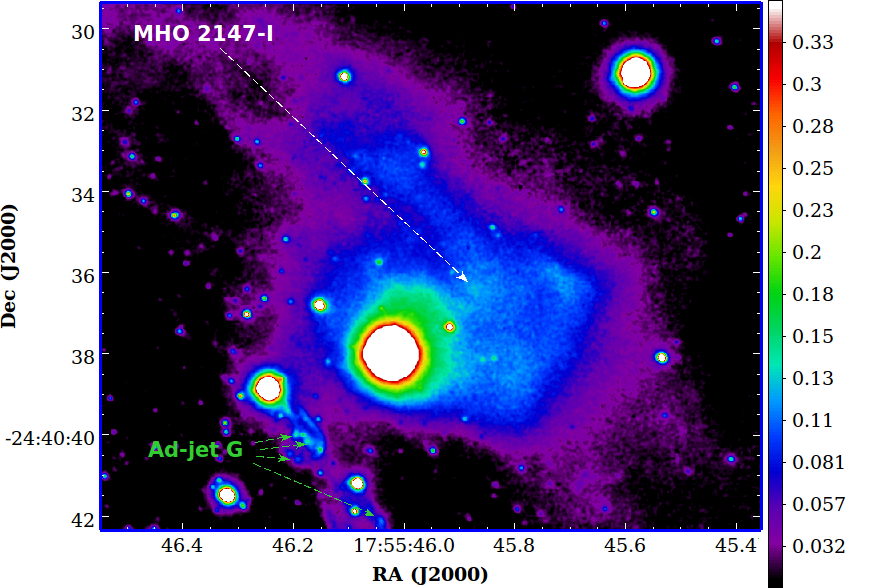}
    \caption{Knots MHO~2147~H (left panel) and I (right panel) in \H2 emission are indicated with dashed-white arrows, and the dashed-green lines mark the knots E, F (left panel) and G (right panel, see also Fig.~\ref{fig_ad-jet_BDG}, bottom panel) which seem to belong to the Ad--jet. The flux scale is shown on the right and is calibrated in units of 10$^{-8}$~Jy.}
    \label{fig_MHO2147_HI}
\end{figure*}


\begin{figure*} 
 \centering
  \includegraphics[width=\columnwidth]{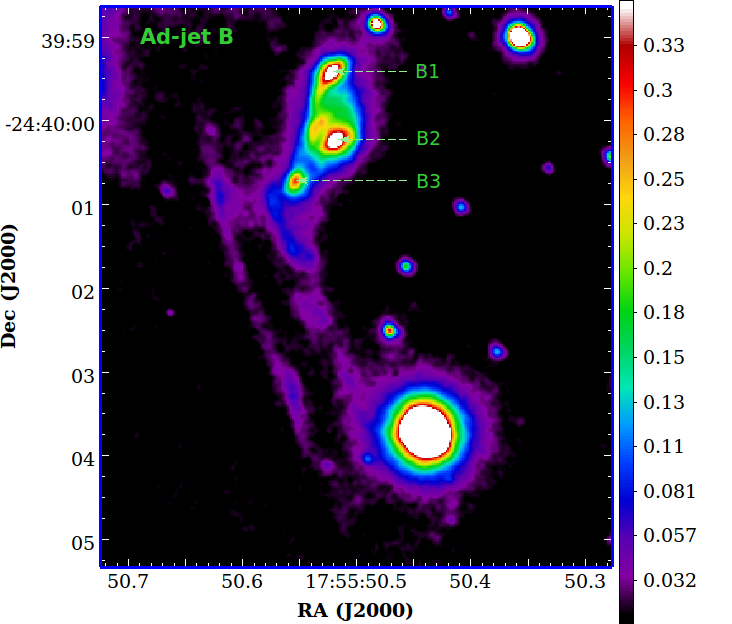}
  \includegraphics[width=\columnwidth]{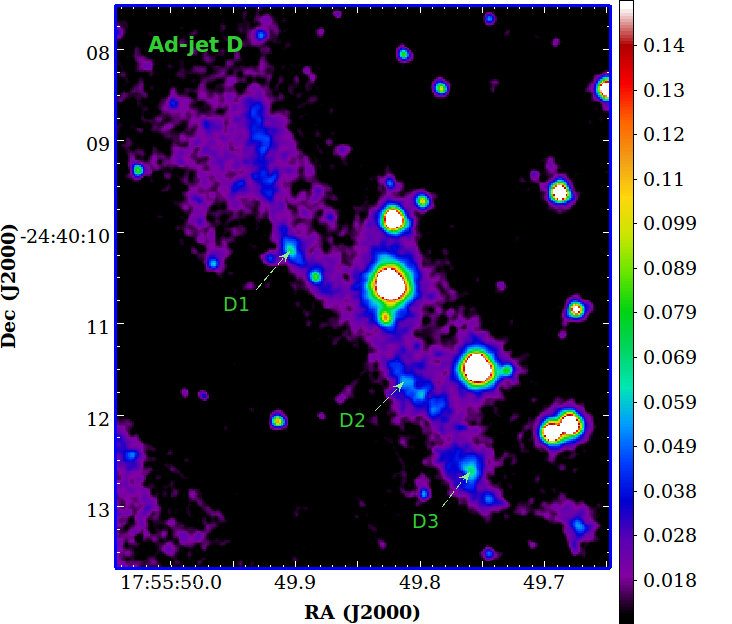}
  \par
  \hspace{0cm}
  \par
  \includegraphics[width=\columnwidth]{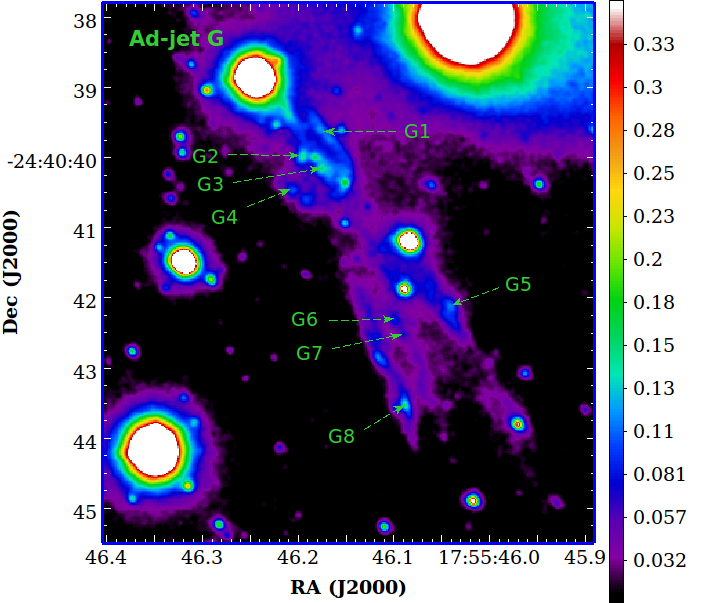}
  \caption{Knots B (upper left panel), D (upper right panel) and G (bottom panel), which seem to belong to the Ad--jet, are indicated with dashed-green arrows. The flux scale is shown to the right and is calibrated in units of 10$^{-8}$~Jy.}
 \label{fig_ad-jet_BDG}
\end{figure*}


\clearpage 
\newpage
\section{Enlarged images of MHO 2148 and coordinates for the \texorpdfstring{\H2}{} knots}
\label{appendix_mho2148}

\begin{figure*} 
 \centering
 \includegraphics[width=0.9\columnwidth]{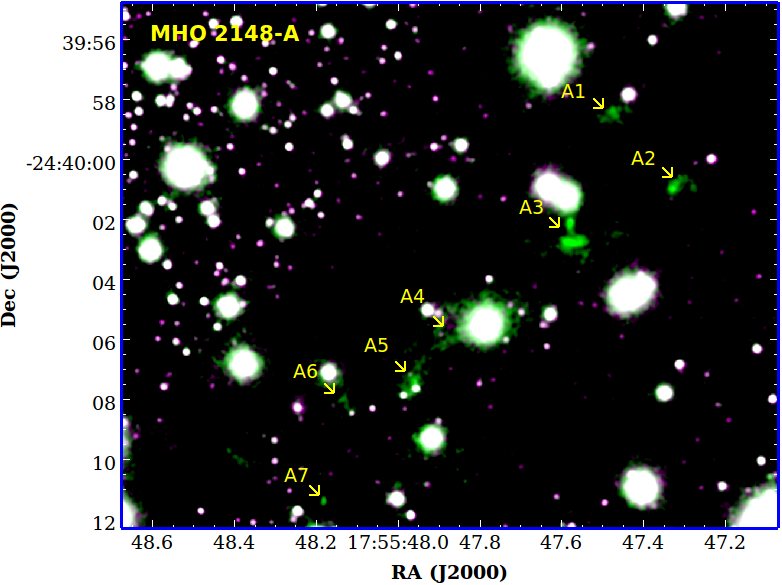}
 \hspace{0.3cm}
 \includegraphics[width=0.9\columnwidth]{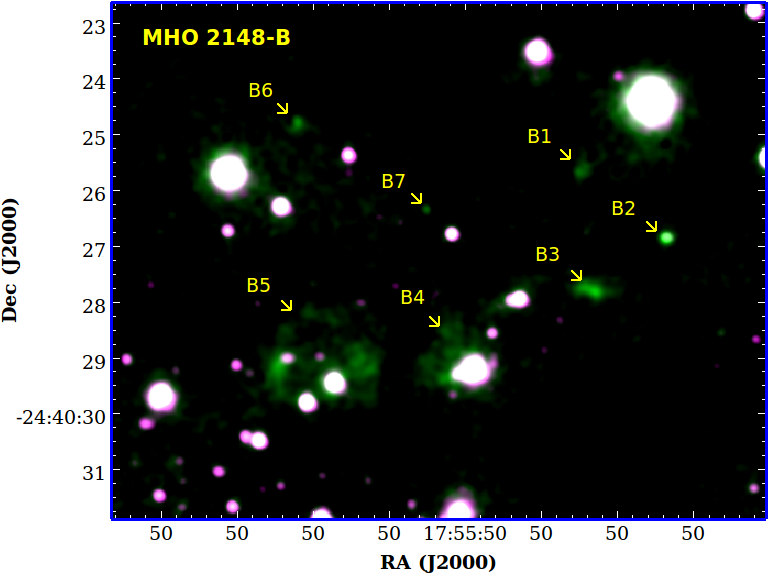}
 \caption{Composite image of the northwest (knots A, left panel) and southeast (knots B, right panel) lobes of MHO~2148. The K-band filter is shown in magenta and the \H2-band filter in green. The \H2 emission structures are marked with yellow arrows.
 To enhance the faint knots a smoothing 6~px ($\sim 0.12\arcsec$) has been applied to the K and \H2 filters. In addition, for the \H2 filter image a logarithmic scale was used.}
 \label{fig_MHO2148_AB}
\end{figure*}


\begin{table*}
    \centering
    \begin{threeparttable}
    \caption{Coordinates for the \H2 knots associated with the MHO~2148 jet.}
    \label{tab_mho2148}
    \begin{tabular}{S[table-text-alignment = center]cc}
    \hline \hline \noalign{\smallskip}
    \multicolumn{1}{c}{\multirow{2}{4em}{Knot ID}} & $\alpha$\,(J2000.0) & $\delta$\,(J2000.0)   \\
         & ($^{h}:^{m}:^{s}$) & 
    (\degr:\arcmin:\arcsec) \\
    \hline \noalign{\smallskip}
    A1  & 17:55:47.48 & -24:39:58.4  \\
    A2  & 17:55:47.31 & -24:40:00.8  \\
    A3  & 17:55:47.57 & -24:40:02.6 \\
    A4\tablefootmark{a}  & 17:55:47.86 & -24:40:05.6  \\
    A5  & 17:55:47.96 & -24:40:07.4 \\
    A6\tablefootmark{a}  & 17:55:48.13 & -24:40:08.5  \\
    A7  & 17:55:48.18 & -24:40:11.4\\
    \hline\noalign{\smallskip}
    B1  & 17:55:49.65 & -24:40:25.6\\
    B2\tablefootmark{b}  & 17:55:49.54 & -24:40:26.8 \\
    B3  & 17:55:49.65 & -24:40:27.6 \\
    B4  & 17:55:49.82 & -24:40:28.9 \\
    B5  & 17:55:49.98 & -24:40:29.1\\
    B6  & 17:55:50.02 & -24:40:24.8 \\
    B7  & 17:55:49.85 & -24:40:26.3 \\
    \hline
    \end{tabular}
    \tablefoot{
    \tablefoottext{a}{Barely detectable.}
    \tablefoottext{b}{The detected \H2 emission superimposes on a likely to be background source in the K-band.}}
    \end{threeparttable}
\end{table*}

\end{appendix}

\end{document}